\documentclass[useAMS,usenatbib]{mn2e}
\usepackage{amsmath,amssymb,mathrsfs,graphicx,float,latexsym,mathtools}
\usepackage{epstopdf,ragged2e}
\usepackage{bm,relsize} 
\usepackage{hyperref}
\usepackage{color}
\usepackage{tikz-cd}
\usepackage{subfigure}
\usepackage{placeins}
\usepackage{tablefootnote}
\newcommand{\Rs}{R_{\star}}
\newcommand{\Ms}{M_{\star}}

\newcommand{\PT}{\Phi^{\text{T}}}
\newcommand{\Mc}{M_{\text{comp}}}
\newcommand{\sm}{\sigma_{\text{max}}}
\newcommand{\td}{e^{i\omega_{\alpha} t}}

\newcommand{\Oo}{\Omega_{\text{orb}}}

\newcommand{\Bn}{B_{15}}
\newcommand{\Mn}{M_{1.4}}
\newcommand{\Rn}{R_{10}}

\newcommand{\Po}{\rho+p}

\newcommand{\Bc}{B_{\star}}

\begin{document}

\title[GR treatment of tidal $g$-mode resonances in NSNS binaries]{General-relativistic treatment of tidal $g$-mode resonances in coalescing binaries of neutron stars. I. Theoretical framework and crust breaking}

\author[Hao-Jui Kuan, Arthur G.~Suvorov, and Kostas D.~Kokkotas]{Hao-Jui Kuan$^{1,2}$,\thanks{E-mail:hao-jui.kuan@uni-tuebingen.de} Arthur G.~Suvorov$^{1}$, and Kostas D.~Kokkotas$^{1}$\\
$^1$Theoretical Astrophysics, Eberhard Karls University of T{\"u}bingen, T{\"u}bingen, D-72076, Germany\\
$^2$Department of Physics, National Tsing Hua University, Hsinchu 300, Taiwan}


\date{Accepted ?. Received ?; in original form ?}

\pagerange{\pageref{firstpage}--\pageref{lastpage}} \pubyear{?}

\maketitle
\label{firstpage}

\begin{abstract}
 
\noindent{During the final stages of a neutron-star binary coalescence, stellar quasi-normal modes can become resonantly excited by tidal fields. If the strain exerted by the excited modes exceeds the extent to which the crust can respond linearly, localised crustal failures may occur. In this work, we re-examine resonant $g$-mode excitations of relativistic neutron stars in the last $\sim$ 10 seconds of an inspiral. We adopt realistic equations of state that pass constraints from GW170817, include 3rd order post-Newtonian terms for the conservation orbital motion, and employ a 2.5 post-Newtonian scheme for gravitational back-reaction. Frequency modulations of the modes due to tidal fields, Lorentz forces, and (slow) rotation are also considered to investigate the maximal strain achievable by resonantly-excited $g$-modes. Depending on the equation of state, degree of stratification, and stellar magnetic field, we find that certain $g$-modes excitations may be able to break the crust some seconds prior to coalescence. 
}

\end{abstract}


\maketitle


\section{Introduction}

Tidal effects in compact binary systems containing at least one neutron star (NS) may be studied by both electromagnetic and gravitational-wave (GW) measurements \citep{Abbott17prl,Abbott17apj12,Abbott17apj13}. Such studies allow one to probe the fundamental properties of the progenitor NSs, such as the equation of state (EOS) \citep{Abbott18prl,Radice18}. In the final stages of a merger, orbital energy and stellar internal energy are redistributed efficiently by tidal force(s) and dissipation. The former excites stellar quasi-normal-modes (QNMs), leading to the transfer of orbital energy into excited modes, thus leaving certain imprints into the orbit evolution [e.g., accelerating coalescence and causing shifts in the GW phase by $f-$mode excitations \citep{Kokkotas:1995xe,Vick:2019cun}]. The latter, resulting from viscosity, damps the excited modes, turning kinetic energy into thermal energy, which can heat up the star to $\sim 10^{8}$ K before merger \citep{Lai:1993di}. In particular, when the tidal-perturbing frequency matches the eigenfrequency of a particular QNM at some point prior to merging, the mode becomes resonantly excited. Mode amplitudes increase rapidly during a period of resonance, possibly straining the crust to the point that quake or fracture events can occur \citep{Horowitz09,Baiko18}. It has been suggested that localised failure events offer a possible mechanism \citep{Tsang:2011ad,pap1,Passamonti:2020fur} to trigger `precursor' events of short gamma-ray bursts \citep{Troja10}. 

In general, identifying the precise conditions under which crustal failure can occur is complicated. In addition to the actual physics of fracturing not being perfectly understood [see Sec.~(2.2) of \cite{Lander19} for a discussion], many factors participate in the straining mechanism, such as: the mass ratio of the binary \citep{Steinhoff16,Hinderer16}, the structure and strength of the stellar magnetic field \citep{Nasiri89,pap1}, the degree of stellar stratification, which affects the $g$-mode spectrum in particular \citep{Xu:2017hqo,Passamonti:2020fur}, the rotation rate \citep{yosh99,Gaertig09,Kruger20}, and the stellar EOS that characterises the internal structure \citep{Lattimer01,Zhou17}.  Electromagnetic byproducts of crustal failures, such as precursor events, may therefore deliver useful information about stellar behavior in the final stages of coalescing binaries.

Moreover, NSs are compact enough that relativistic effects are not negligible in these last stage. For instance, QNM eigenfrequencies can differ from their Newtonian counterparts by $\gtrsim 10\%$ \citep{Chan2014}, which, if unaccounted for, results in errors in the estimation of parameters that allow for resonances to happen at certain times. Building on previous studies \citep{Tsang:2011ad,pap1,Passamonti:2020fur}, we introduce a general-relativistic framework in this study that aims to (at least phenomenonologically) incorporate each of the above elements to better understand the connection between resonantly excited modes and crust yielding. This is the first of two papers in a series, where the framework is detailed. In a forthcoming paper (paper II), the aforementioned electromagnetic byproducts are examined in detail and compared with the results obtained herein.

At the non-rotating level\footnote{Because mature NSs as part of binary systems are expected to be slowly rotating \citep{Koch92,Bildsten92}, the inertial-frame and rotating-frame frequencies of the modes roughly coincide. For rapidly rotating stars however, $r$- and even $f$-mode frequencies can be comparable with the frequency of tidal driving $\sim$ seconds before merger \citep{Pnigouras19,pap1}; see Sec. \ref{sec.V.C} for a discussion on rotational corrections.}, the QNMs of NSs can be generally resolved into $p$-, $f$-, $w$-, and $g$-modes. Since the (rotating-frame) frequencies of the stellar $g$-modes, which are QNMs restored by buoyancy, are typically in the hundreds of Hz \citep{McDermott83,Finn87,McDermott1990,Xu:2017hqo}, these modes are generally thought to lie in the sweet spot of the precursor scenario (that is, they match well with the expected driving frequency at the time when precursors are observed relative to the main burst; cf. paper II). We derive empirical relations for EOS- and stratification-related effects on the $g$-mode eigenfrequencies. Shifts in the spectra due to magnetic fields (Section \ref{sec.V.A}), tidal fields (Section \ref{sec.V.B}), and rotation (Section \ref{sec.V.C}), are also considered. In principle interface modes \citep{McDermott1985,McDermott1988,Piro05} and shear modes \citep{Schumaker1983,Sotani:2006at,Vavoulidis:2007cs,Sotani16} could potentially be responsible for precursors as well \citep{Passamonti:2020fur}. However, the stars considered here have neither phase transitions that result in density jumps inside the star, nor a solid crust separated from the fluid core, hence those modes are absent.

This paper is organised as follows.
We write down the equations of stellar structure relevant for the EOS considered here (Sec.~\ref{sec.II.A}). 
Numerical details concerning the calculation of QNMs, with a specific focus on the $g$-modes, is also given (Sec.~\ref{sec.II.B}).
The orbital dynamics and relevant assumptions concerning the binary itself are given in Sec.~\ref{sec.III}.
Magnetic fields with hybrid structure on a relativistic star is derived in Sec.~\ref{sec.IV}. In Sec.~\ref{sec.V} we study mode modulations by tidal, magnetic, and centrifugal forces and the strain generated by excited modes is investigated in Sec.~\ref{sec.VI}. A discussion is offered in Sec.~\ref{sec.VII}.

Except where stated otherwise, quantities are expressed in natural units with $c=G=1$, Greek letters denote four-dimensional spacetime indices with an exception of $\alpha$, which denotes the quantum number of eigenmodes, and Latin indices refer to the spatial 3-components.
We adopt the following notation for compactness throughout:
$\Bn=\Bc/(10^{15}\text{ G})$, $\Mn=M/(1.4\text{ }M_{\odot})$, and $\Rn=R/(10\text{ km})$, where $\Bc$ is the characteristic magnetic field strength that will be introduced in Sec.~\ref{sec.IV.A}.

\section{Stellar structure} \label{sec.II}
We consider a static, spherically symmetric line element
\begin{equation} \label{eq:statsphsym}
	ds^2 = - e^{2 \Phi(r)} dt^2 + e^{2 \lambda(r)} dr^2 + r^2( d\theta^{2}+\sin^{2}\theta^{2}d\phi^{2}) ,
\end{equation}
where $(t,r,\theta,\phi)$ are the usual Schwarzschild coordinates, and $\Phi$ and $\lambda$ are 
functions of $r$ only. The Einstein equations 
\begin{equation} \label{eq:einstein}
	G_{\mu \nu} = 8 \pi T_{\mu \nu},
\end{equation}
for the stress-energy tensor associated with a single, perfect fluid, 
\begin{equation} \label{eq:stresstensor}
	T^{\mu \nu} = \left( \rho + p \right) u^{\mu} u^{\nu} + p g^{\mu \nu},
\end{equation}
describe the structure of a static, non-rotating star.
Here $\rho$ is the energy-density, $p$ is the stellar pressure, $g_{\mu\nu}$ is the metric tensor
defined in \eqref{eq:statsphsym}, and $u^{\mu}=e^{-\Phi} \delta^{\mu}_{0} \partial_t$ is the 4-velocity of a generic fluid element (rotational corrections to the stellar structure are considered in Sec.~\ref{sec.V.C}).
The metric function $\lambda$ is related to the mass distribution function $m(r)$, 
which yields the mass inside the circumferential radius $r$, through
\begin{equation}
	e^{-2 \lambda} = 1 - \frac {2 m(r)} {r}.
\end{equation}
The conservation law,
\begin{equation} \label{eq:euler}
	0 = \nabla^{\mu} T_{\mu \nu},
\end{equation} 
relates the functions $\rho(r)$, and $p(r)$ to the metric variables, and forms the following system 
\begin{subequations}
	\begin{align}
		\frac {d \Phi} { d r} = \frac {1} {p(r) + \rho(r)} \frac {d p} {d r}, \label{eq:tov1}
	\end{align}
	\begin{align}
		\frac {d m} {d r} = 4 \pi r^2 \rho(r), \label{eq:tov2}
	\end{align}
and
	\begin{align}
		\frac {d p} {d r} = - \frac {\left[ \rho(r) + p(r) \right] \left[ m(r) + 4 \pi r^3 p(r) \right]} { r^2 \left[ 1 - \frac {2 m(r)} {r} \right]}.\label{eq:tov3}
	\end{align}
\end{subequations}
The star's radius $\Rs$ and mass $\Ms$ are defined by the boundary conditions $p(\Rs) = 0$ and  $m(\Rs) = \Ms$, respectively.
Outside of the star, where $p = \rho = 0$, the metric \eqref{eq:statsphsym} reduces to the Schwarzschild metric of mass $\Ms$.

\subsection{Equation of state}\label{sec.II.A}

We consider piecewise-polytropic approximations \citep{Read:2008iy} to three different realistic EOS, namely the APR4, SLy, and WFF families. We choose these models because they are sufficiently soft to be compatible with the tidal deformability measured in GW170817 \citep{Abbott18prl}. The aforementioned EOS are all barotropic [i.e., $p = p(\rho)$], which is a reasonable approximation for mature systems older than the relevant electroweak and diffusion timescales \citep{mast15} where thermal fluxes are likely to be negligible, and the buoyancy comes primarily from composition gradients inside the star. Note, however, that tidal heating is expected to be able to raise the temperature of the (still relatively cold) NS crust to $\sim10^{8}$ K prior to merger \citep{Lai:1993di}. Thermal gradients may therefore become important at late times \citep{Bauswein10}.

\begin{figure}
	\centering
	\includegraphics[width=0.45\textwidth]{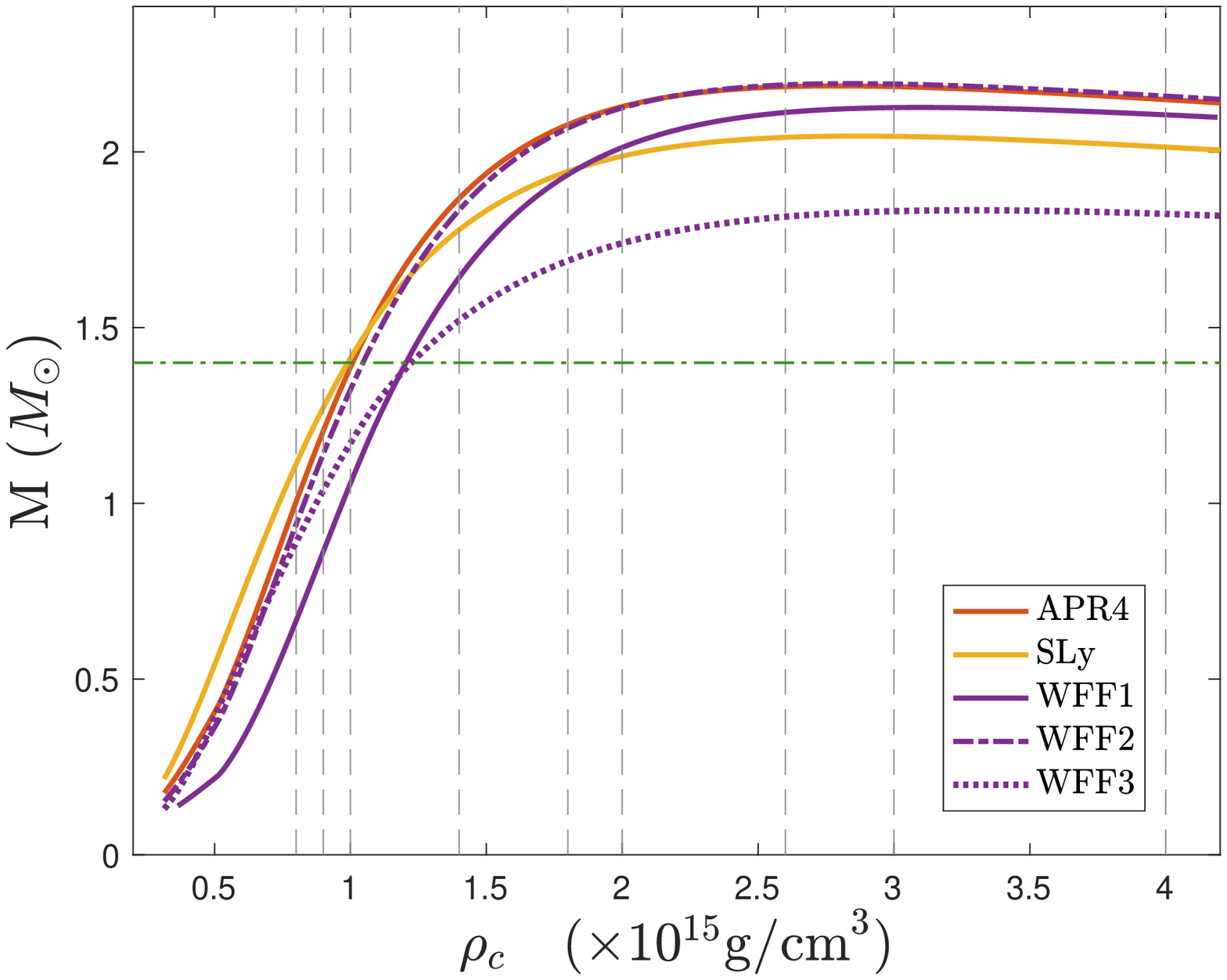}
	\includegraphics[width=0.45\textwidth]{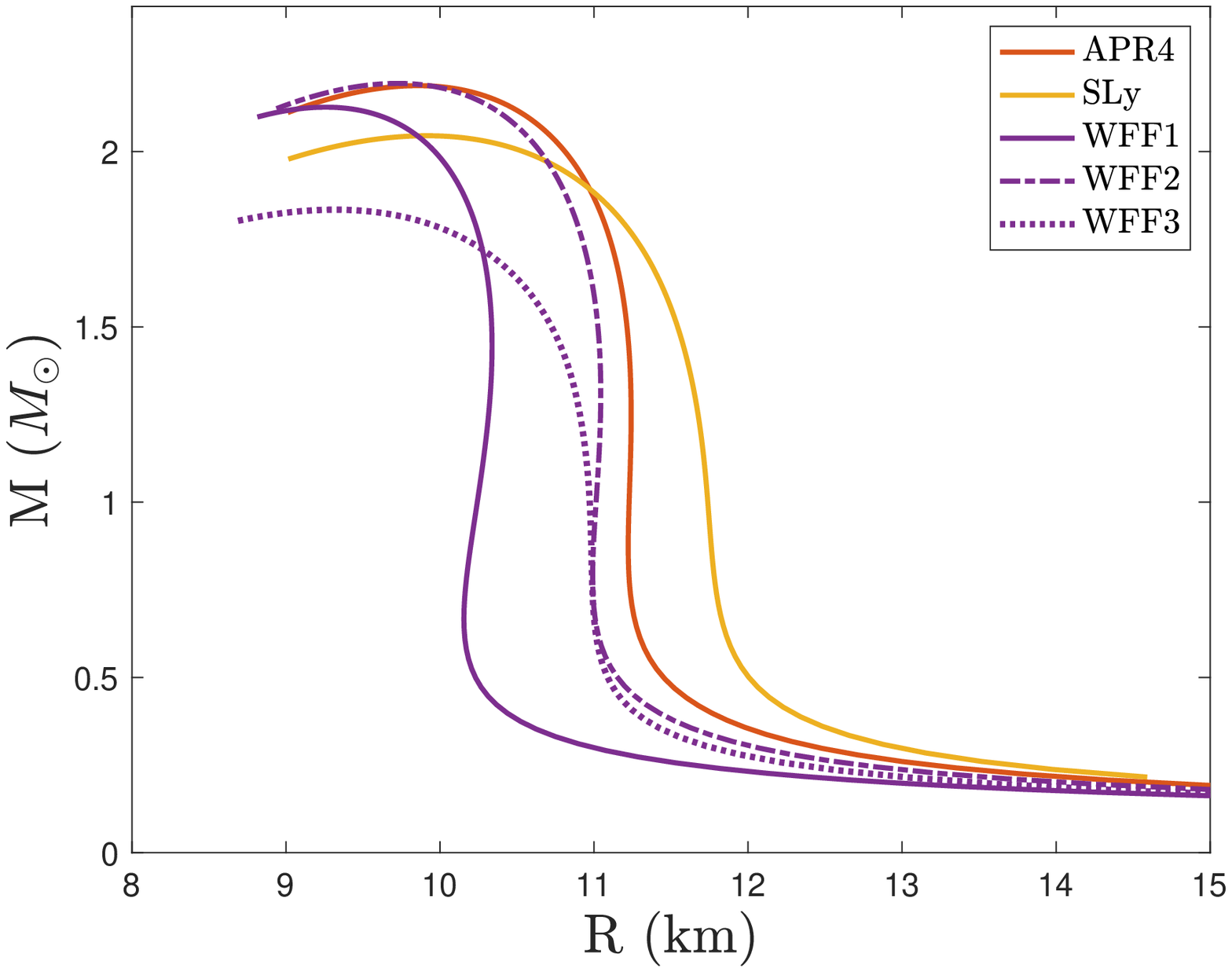}
	\caption{ Mass of non-rotating equilibrium models constructed with various EOSs surviving the constraints of GW170817 (APR4, SLy, WFF1-3) as functions of central density (top panel) and stellar radius (bottom panel).
	We choose several models for each EOS to be studied. These models include those represented by the intersection of each EOS curve with the gray dashed lines, and those having masses of $1.4M_{\odot}$ (the intersection of each EOS curve with the green dashed line).
	}
	\label{fig:eosmodels}
\end{figure}

While each EOS considered here assumes that the stars consist of $npe\mu$ nuclear matter, the many-body problem is handled differently:
\begin{enumerate}
	\item WFF families are obtained using variational methods applied to nucleon Hamiltonians, that contain pieces of two-body and three-body interactions.
	More precisely, different two- and three-nucleon potentials are used to model the bulk matter [see  \cite{WFF} for details].
	\item SLy is derived from the Skyrme effective nucleon-nucleon interaction \citep{SLy}, consistent with WFF2 in the regime where the baryon density exceeds the nuclear value $n_{0}=0.16$ fm$^{-3}$.
	\item APR4 is derived by variational chain summation methods \citep{APR} adopting a two-nucleon interaction \citep{Wiringa:1994wb} that accounts for Lorentz boost corrections not used in WFF1.
\end{enumerate}
Masses $\Ms$ of the stars constructed with these EOS, as functions of central density, are shown in the upper panel of Figure \ref{fig:eosmodels}.
The intersections of each curves with the dashed lines (both green and gray) mark the models that we choose for later analysis.
The mass-density relations [bottom panel of Fig.~\ref{fig:eosmodels}] tells us that the SLy EOS is the stiffest one and the WFF1 is the softest one.

\subsection{$g-$modes}\label{sec.II.B}

Oscillatory patterns of motion in a star can be resolved into QNMs with complex eigenfrequency $\omega_{\alpha}$, where $\alpha$ denotes the ensemble of quantum numbers $\alpha = (nlm)$ for overtone number $n$, and spherical harmonic indices $l$ and $m$.
The oscillating frequency is the real part of $\omega_{\alpha}$, while the reciprocal of the imaginary part defines the damping rate due to radiation-reaction.
In general, QNMs are categorised according to the nature of the restoring force. For example, $p-$modes are supported by pressure, while $g$-modes are supported by buoyancy.
We focus on $g$-modes in the present work because they are more likely to become resonant in the last $\sim 10$ s of inspirals for slowly-rotating NSNS mergers (though see Footnote 1) since their (fundamental) frequencies are typically smaller than a few hundred Hz \citep{Lai:1993di,Kokkotas:1995xe,Kantor14,Andersson18}.

Inhomogeneities in composition and/or temperature give rise to stellar stratification and buoyancy as gravity tends to smooth out these gradients \citep{McDermott83,Finn87,Strohmayer91,Reis92}. 
Explicitly, stratification prevents the Schwarzschild discriminant,
\begin{align}
	A \coloneqq e^{-\lambda}\frac{dp}{dr}\frac{1}{p}\bigg( \frac{1}{\gamma}
	-\frac{1}{\Gamma} \bigg),
\end{align}
from vanishing, where 
\begin{align}
	\gamma=\frac{\Po}{p}\frac{dp}{d\rho}
\end{align}
is the adiabatic index associated with the equilibrium star described in Sec.~\ref{sec.II}. The parameter $\Gamma$ represents the adiabatic index of the \emph{perturbation}, which need not match that of the background for non-isentropic perturbations \citep{Lockitch01}. Aiming to provide a proof-of-principle framework in the present work, we introduce a phenomenological parameter $\delta$ which encapsulates the departure from isentropicity through 
\begin{align}
	\Gamma=\gamma(1+\delta).
\end{align}
In principle, one could determine $\delta$ from first principles by calculating the sound speed and the determinant of the Brunt-V{\"a}is{\"a}l{\"a} frequency from the nuclear Hamiltonian together with the Gibbs equation describing the evolution of the chemical composition \citep{Reis92,Lai:1993di}. However, here we approximate the EOS as barotropic, i.e., $p=p(\rho)$, which erases the compositional information in practice. The composition gradient, which offers buoyancy for $g$-modes, is therefore, strictly speaking, absent. The artificial parameter $\Gamma$ is used as a proxy for perturbation-induced changes in the chemical potentials resulting from a  \emph{non-adiabatic} perturbation. In addition, as shown by \cite{Reisenegger:2001js}, NSs are in general stably stratified due to the interior equilibrium composition gradient, implying $A<0$ inside the star (i.e.~$\gamma<\Gamma$). We thus consider positive $\delta$ hereafter.

\begin{figure}
	\centering
	\includegraphics[width=0.5\textwidth]{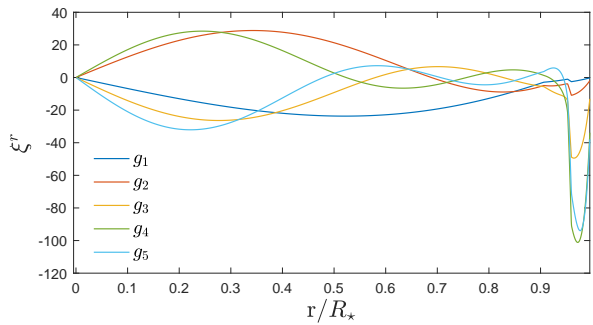}
	\caption{ Normalized radial displacement $\xi^r$ for the $n=1-5$, $l=2$, $g$-modes of the star with EOS SLy, $\Ms=1.27M_{\odot}$, and $\delta=0.005$.		
	}
	\label{fig:gmodes}
\end{figure}

The numerical calculation of the complex $g-$mode frequencies is known to be difficult because $|\text{Im}(\omega_\alpha)| \ll |\text{Re}(\omega_\alpha)|$, meaning that high precision is necessary to prevent errors in the real parts contaminating the imaginary parts, as discussed by \cite{Finn:1986}.
Searching for low-frequency $g$-modes and their respective eigenfunctions entails a delicate separation of the ingoing- and outgoing-waves at spatial infinity, so that one can impose the purely outgoing boundary condition\footnote{Solutions to the perturbation equations include those of purely ingoing- and outgoing-waves, and even the hybrid waves. However, only those pulsate energy outward, i.e.~purely outgoing ones, are physically realized \citep{Price:1969}. }.
Techniques based on \textit{phase integrals} have been proven to be adequate for this purpose \citep{Andersson:1995wu}.
On top of that, the minute displacements of $g-$modes, which is translated from eigenfunctions, make the differential system of the eigenproblem put forward by \cite{Lindblom:1983,Detweiler:1985} inappropriate for solving $g-$modes.
The issue for these long-lived modes (due to their small imaginary components) has been addressed in \cite{Kruger:2014pva} by solving a slightly different set of differential equations [see also \cite{Finn:1986}].
In this work we adopt the combined algorithm of \cite{Andersson:1995wu} and \cite{Kruger:2014pva} 
to compute $g-$modes.
Our code is capable of determining the real parts of mode frequencies to within a tolerance of $\sim 10$ Hz.
Shown in Fig.~\ref{fig:gmodes} are the radial displacement $\xi^r$ for the first five $g$-modes of a particular star.

There is a universal relation for the frequencies of $f-$modes over a plethora of EOS for non-rotating NSs \citep{Andersson:1997rn}  and for (rapidly-)rotating NSs \citep{Kruger20}, which can be used to infer the eigenfrequencies of $f-$modes by the mean density of the background star and vice versa \citep{Andersson96,Lau10,Volkel2021}. Similar relations have been found for other modes also \citep{Andersson:1997rn,Kokkotas01,Tsui05}.
In particular, though restricted to polytropic EOS, \cite{Xu:2017hqo} found universal relations between the frequencies of $l=2$, $g_1$-modes and the strength of stratification (i.e., $\delta$ in our notation). 
Similar to \cite{Xu:2017hqo}, we find that the real parts of the eigenfrequencies are well-described, as a function of $\delta$, by
\begin{align}\label{eq:fitomega2del}
	\text{Re}[\omega_{g_1}]  \propto \bigg(\frac{M^{\varsigma}}{R^{1+\varsigma}}\bigg)\sqrt{\delta},
\end{align}
where the fitting constants $\varsigma$ for each EOS are listed in Table \ref{tab:constant_a}.
For the polytropic EOS studied in \cite{Xu:2017hqo}, the corresponding constant is $\varsigma=0.5$ --- it is lower for realistic EOS.

\begin{table}
	\centering
	\caption{Fitting index $\varsigma$, defined in Eq.~\eqref{eq:fitomega2del}, for leading-order $g-$mode frequencies as a function of $\delta$.}
	\begin{tabular}{|c|c|c|c|c|c|}
		\hline
		EOS & APR4 & SLy & WFF1 & WFF2 & WFF3 \\
		\hline
		$\varsigma$ & 0.231 & 0.378 & 0.148 & 0.212 & 0.465 \\
		\hline
	\end{tabular}
	\label{tab:constant_a}
\end{table}

\section{Binary Evolution}\label{sec.III}
We consider a close NSNS binary system with constituent masses $\Ms$ and $M_\text{comp}$ for the primary and companion, respectively.
The orbit is assumed to lie on the equatorial plane\footnote{In close binaries, tidal interaction aligns the stellar spins with the orbital angular momentum rapidly \citep{Hut81,Zahn08}. Therefore, it is most likely that stellar spins are almost aligned with the orbital one in the late stage of inspiralling.}.
Each star perceives the other as a point mass to leading order, and thus we treat the companion as a point mass in the evolutionary code\footnote{Higher-order and finite-size effects have been looked at through the second-order gravitational self-force method \citep{Pound12} and, independently, by post-Newtonian theory \citep{Bini12}, where the leading correction comes beyond 5 PN.}. The relevant Hamiltonian consists of four parts \citep{Alexander:1987zz,Kokkotas:1995xe}:
\begin{enumerate}
	\item The conservative orbital dynamics, for which we include up to 3rd order post-Newtonian (PN) corrections via the effective one-body formalism [since the equations are lengthy, we refer the reader to Eq.~(4.28) of \cite{Damour:1999cr}; see also \cite{Buonanno:1998gg}].
	\item Leading order GW dissipation of respective equatorial motions, which first appear at 2.5 PN order, is encapsulated by \citep{schaefer90}
	\begin{align}\label{eq:Hreact}
		H_{\text{react}} = \frac{2}{5}\bigg(  p_{i}p_{j}- \frac{\Ms^{2}\Mc^{2}}{\Ms+\Mc}  \frac{x_{i}x_{j}}{a^{3}} \bigg)  \frac{d^{3}}{dt^{3}}\bigg[ x_{i}x_{j}-\frac{a^{2}}{3}\delta_{ij} \bigg],
	\end{align}
	where $x_{i} = (a\cos\phi, a\sin\phi)$ for $i= 1,2$ denote the spatial coordinates of the companion (note the lack of dependence on $\theta$ because we consider equatorial motion), $p_i$ are the associated momenta, and $a$ is the distance between the companion and the center of the primary $(r=0)$.
	
	\item The gravitational energy that the primary absorbs via the tidal field,
	\begin{align}
		H_{\text{tid}} = \int_{\text{primary}}\PT\delta\rho(\mathbf{x},t) \sqrt{-g}d^3x,
	\end{align}
	where $\Phi^{T}$ denotes the tidal potential as experienced by the primary, $\delta \rho$ is the physical variation in density, and the integral is taken over the volume of the primary. We note that the tidal potential $\PT$ is not to be confused with the metric function $\Phi(r)$.
	In true general relativity, the tidal potential $\PT$ (which is promoted to a tensor) has both electric and magnetic components \citep{Landry15,Poisson17,Poisson20} [see also \cite{Vines:2010ca,Ferrari12,Steinhoff16} for the formalism of 1 PN interaction]. However, the PN tidal response of NSs and the influence on the orbit evolution are insignificant except only the last few second of the insprial $(\gg 100 \text{ Hz}$). Therefore, we will neglect the PN tidal interaction and consider a Newtonian approximation here.
	As such, $\Phi^{T}$ admits a multipole expansion of the form \citep{Press77} [see also \cite{zahn77} for the case of eccentric binaries] 
	\begin{equation} \label{eq:tidalpot}
		\PT= - \frac {\Mc} {a} \left[ 1 + \sum_{l = 2} \left( \frac {r} {a} \right)^{l} P^{l}_{0}  \left(\cos \tilde{\phi} \sin \theta \right) \right],
	\end{equation}
which depends on the difference $\tilde{\phi} = \phi - \phi_{c}$ between $\phi$ and the angular position of the secondary star, $\phi_{c}$, as measured from the perihelion of the orbit.
	In general one needs to sum over the multipolar components of $\PT$ to complete the tidal Hamiltonian, though we specialize our attention to the $l=m=2$ component of $\PT$, which is the leading order term of the potential most relevant for tidally-forced oscillations \citep{zahn77,Willems03}.
	The tidal force associated with this component perturbs the primary with frequency two times the orbital frequency, $2\Omega_{\text{orb}}$.
	In addition, $\delta\rho(\mathbf{x},t)$ is induced by the small-amplitude motion $\xi$ on the star, 
	\begin{align}\label{eq:sumoverxi}
		\xi(\mathbf{x},t) = \sum_{\alpha}q_{\alpha} (t) \xi_{\alpha}(\mathbf{x},t),
	\end{align}
	which we have decomposed into modes $\xi_\alpha$ with amplitude $q_{\alpha}$ (see Sec.~\ref{sec.III.A} for details).
	Each $\xi_{\alpha}$, having time dependence $\td$, is a solution to the eigenproblem 
	\begin{align}\label{eq:operat}
		\mathcal{V} \xi_{\alpha} = \omega_{\alpha}^{2} \mathcal{T} \xi_{\alpha},
	\end{align}
	where $\mathcal{V}$ and $\mathcal{T}$ are appropriate potential and kinetic operators \citep{Fuller:2020}. The detailed form of these potentials is crucial as one attempts to identify the impact of any perturbing forces in the problem, but is not important in building up the Hamiltonian itself, so we postpone their explicit definition until Sec.~\ref{sec.V.B}.
	\item Pulsations on the primary, which are described by a 
	harmonic-oscillators-type Hamiltonian
	\begin{align}\label{eq:osc}
		H_{\text{osc}}=\frac{1}{2}\sum_{\alpha}
		\bigg( \frac{p_{\alpha}\bar{p}_{\alpha}}{\Ms\Rs^{2}}+{\Ms\Rs^{2}}\omega_{\alpha}^{2}q_{\alpha}\bar{q}_{\alpha} \bigg) + \text{H.c.},
	\end{align}
	 which is normalized according to
	\begin{align}
		\int_{\text{primary}}\sqrt{-g}d^{3}x e^{-2\Phi} (\Po) (\xi_{\alpha})^{\mu} (\bar{\xi}_{\alpha'})_{\mu} = \Ms\Rs^{2}\delta_{\alpha\alpha'}
		\label{eq:normalization}
	\end{align}
	for each QNM eigenfunction\footnote{Eigenfunctions of QNMs, $\xi_{\alpha}^{\mu}$, in GR are not strictly orthogonal to each other for the coupling between the material motion to the gravitational radiation field, which extends to infinity, destroys the self-adjointness of the eigenvalue problem by a non-vanishing surface integral term from the perturbations in the spacetime [see, e.g., Eq.~(2.4) in \cite{Fried75}; for the Cowling approximation case, see the last two terms of Eq.~(16) in \cite{Detweiler73}]. 
		Nonetheless, that term is small for $g-$modes, whose dissipation timescales are extremely long.
		The omission of the surface integral term hence justifies the implementation of the normalization \eqref{eq:normalization}, which looks similar to the Newtonian case used in, e.g., \cite{Tsang:2011ad}. 
	}. Here $p_{\alpha}$ are the canonical momenta associated with $q_{\alpha}$, and the overhead bar denotes complex conjugation.
	Note that the momenta with Latin index are spatial ones, while those labeled by $\alpha$ are for pulsations.
	The Hermitian conjugate in Eq.~\eqref{eq:osc} comes from the dual appearance of modes with eigenfrequency $-\bar{\omega}$ (see Sec.~\ref{sec.III.A} for details).
	However, these are not the classic oscillators in the sense that dissipation rate of QNMs is not determined solely by the imaginary part of the eigenfrequencies, since the eigenfunctions are not real.
	
\end{enumerate}

In summary, we work with the Hamiltonian
\begin{align}\label{eq:ham}
	H(t)=\big(H_{\mathrm{orb}} + H_{\mathrm{reac}} + H_{\mathrm{osc}} + H_{\mathrm{tid}} \big) (t).
\end{align}
The orbital dynamics are then determined by numerically solving Hamilton’s equations,
\begin{subequations}
	\begin{align}\label{eq:eomosc}
		\frac{dp_{\alpha}}{dt} = -\frac{\partial H(t)}{\partial q_{\alpha}}, \;\;
		\frac{dq_{\alpha}}{dt} = \frac{\partial H(t)}{\partial p_{\alpha}}, \\
		\frac{d\bar{p}_{\alpha}}{dt} = -\frac{\partial H(t)}{\partial \bar{q}_{\alpha}}, \;\;
		\frac{d\bar{q}_{\alpha}}{dt} = \frac{\partial H(t)}{\partial \bar{p}_{\alpha}},
	\end{align}
and
	\begin{align}\label{eq:eomorb}
		\frac{dp_{i}}{dt} = -\frac{\partial H(t)}{\partial x_{i}}, \;\;
		\frac{dx_{i}}{dt} = \frac{\partial H(t)}{\partial p_{i}},
	\end{align}
\end{subequations}
where we recall that $x_i$ and $p_i$ are defined in the sentence below Eq.~\eqref{eq:Hreact}.

The evolution is carried out up to the point that the orbital instability kicks in, which happens at $a \lesssim 3q^{1/3}\Rs$  \citep{Lai93}, where $q$ is the mass ratio $\Mc/M$ of the binary. Although this point need not coincide with the separation where NSs merge, the difference is small \citep{Lai94b} and we effectively assume that mergers occur at $a \lesssim 3q^{1/3}\Rs$ \citep{Lai94a,Ho99,pap1}.

\subsection{Tidal resonances}\label{sec.III.A}

For a spherically-symmetric (equilibrium) star, the components of the eigenfunction $\xi_\alpha$ can be expressed in terms of radial ($W_{nl}$) and tangential ($V_{nl}$) components, viz. 
\begin{align}\label{eq:xi}
	&\xi_{\alpha}^{r} = r^{l-1} e^{-\lambda} W_{nl}(r) Y_{lm}\td, \nonumber\\
	&\xi_{\alpha}^{\theta} = -r^{l-2}V_{nl}(r)\partial_{\theta}Y_{lm}\td, \nonumber\\
	&\xi_{\alpha}^{\phi}= -r^{l}(r\sin\theta)^{-2}V_{nl}(r)\partial_{\phi}Y_{lm}\td,
\end{align}
and $\xi_{\alpha}^{t}=0$ \citep{Thorne:1967,Lindblom:1983,Detweiler:1985}.
In addition, the metric perturbed by (even parity) QNMs can be expressed in the Regge-Wheeler gauge\footnote{
	Strictly speaking, this gauge assumes a fixed $(l,m)$ and mode parity [cf.~Eq.~(A9) and (A11) in \cite{Thorne:1967}], and so performing a summation, as we do in \eqref{eq:sumoverxi}, is actually mixing gauges in a formally incorrect way.
	Fortunately, \cite{Price:1969} have shown that one can simply superpose the QNMs using whatever gauge for each one.
} as
\begin{align}\label{eq:fullmetric}
	ds^{2} =& ds_{\text{eq}}^2-e^{2\Phi}  r^{l}H^{0}_{nl}Y_{lm}\td   dt^{2} -2i\omega_{\alpha} r^{l+1} H^{1}_{nl}Y_{lm}\td dtdr \nonumber\\
	& - e^{2\lambda} r^{l}H^{0}_{nl}Y_{lm}\td  dr^{2}
	 -  r^{l+2}K_{nl}Y_{lm}\td d\Omega^{2},
\end{align}
where $ds_{\text{eq}}^2$ is the line element of the equilibrium \eqref{eq:statsphsym}, and $H^{0}_{nl}$, $H^{1}_{nl}$, and $K_{nl}$ characterize the metric perturbations.

Two modes whose eigenfrequencies have real parts with opposite sign but share the same imaginary parts appear in pairs \citep{Andersson:1997eq}, and their eigenfunctions are complex conjugate to each other.
The normalization \eqref{eq:normalization} is satisfied for these dual modes, thus the Hermitian conjugate part in Eq.~\eqref{eq:osc} attributes to them.
The change in the (Eulerian) density induced by pulsations is therefore
\begin{align}
	\delta\rho(\boldsymbol{x},t) = \sum_{\alpha}  \delta\rho(\boldsymbol{x},\omega_{\alpha}) \td+ \text{H.c.},
\end{align}
where the contribution of a particular mode, accompanying a Hermitian conjugate term due to the dual mode, is
\begin{align}
	\delta\rho(\mathbf{x},\omega_{\alpha}) =& q_{\alpha}  \bigg[-e^{-\Phi}\nabla_{i} \bigg( (\Po)e^{\Phi}\bxi^{i}_{\alpha} e^{-i\omega_{\alpha} t}\bigg) \nonumber\\
	&+ \bigg( \frac{H^{0}_{nl}}{2}+K_{nl} \bigg)(\Po) Y_{lm}\bigg]
\end{align}
to first order in the perturbation terms [cf.~Eq.~(8a) in \cite{Detweiler73}].
The boldface symbol denotes the spatial part of a 4-vector and the divergence is taken with respect to the 3-geometry of the metric \eqref{eq:fullmetric} at a constant time $t$.
 The physical perturbation in density induced by a pair of modes reads
\begin{align}
	\delta\rho(\mathbf{x},t) = 2\mathbf{\text{Re}}[\overline{\delta\rho}(\mathbf{x},\omega)].
\end{align}
We use the complex conjugate $\overline{\delta\rho}$ in the bracket to maintain coherence with later use. 
The factor $2$ comes from the fact that the modes appear in pairs with frequency of $\omega$ and $-\bar{\omega}$.

Substituting $\delta \rho$ and integrating by parts, the tidal Hamiltonian can be expressed as
\begin{align}\label{eq:Htid}
	H_{\text{tid}}=&2 \int_{\text{primary}} \sqrt{-g}d^3x (\Po) \text{Re}\big[ \overline{\delta\rho}(\mathbf{x},\omega) \PT  \big]  \nonumber\\
	=&\int_{\text{primary}} \sqrt{-g}d^3x (\Po) \text{Re}\big[ (\bar{H}^{0}_{nl}+2\bar{K}_{nl})\bar{Y}_{lm}\PT \big] \nonumber\\
	& -\frac{2\Ms\Mc}{a\Rs}\sum_{\alpha} W_{lm}  \bigg(\frac{\Rs}{a}\bigg)^{l}  \text{Re}[\bar{q}_{\alpha}Q_{\alpha}e^{-im\phi_{c}}],
\end{align} 
containing a term resulting from the spacetime distortion, which does not have a Newtonian analogy.
In Eq.~\eqref{eq:Htid}, $W_{lm}$ is given by
\begin{align}
	W_{lm}=&(-)^{(l+m)/2}\left[{4\pi\over 2l+1}(l+m)!(l-m)!\right]^{1/2} \nonumber\\
	&\times \left[2^l\left({l+m\over 2}\right)!\left({l-m\over 2}\right)!\right]^{-1},
\end{align}
where $(-)^{k}=(-1)^{k}$ if $k$ is an integer, but equals zero otherwise. 
The relativistic ``overlap integral'', defined as \citep{Press77}
\begin{align} \label{eq:tidaloverlap}
	Q_{nl} = \frac{1}{\Ms\Rs^{l}} \int_{\text{primary}}\sqrt{-g} d^{3}x(\Po) \bar{\xi}_{nll}^{\mu} \nabla_{\mu} (r^{l} Y_{ll}),
\end{align}
is a complex, \emph{dimensionless} number which measures the tidal coupling strength of the mode.
The tidal overlap integral for the predominant effects ($l=m=2$ component of $\PT$) reads\footnote{
	Note that this expression differs from that used by \cite{yosh99}. These latter authors ignored the pressure contribution in addition to the inertial mass in their expression.
}
\begin{align}
Q_{n2} &=\frac{1}{\Ms\Rs^{2}}\int_{\text{primary}} e^{\Phi+\lambda} (\Po)  
\bar{\xi}_{n22}^{\mu} \nabla_{\mu}(r^{2}  Y_{22}) r^{2}d^{3}x.
\label{eq:overlap}
\end{align}

For a binary system, the tidal force has the frequency of $2 \Omega_{\text{orb}}$ \citep{zahn77}, thus pulsation modes, with eigenfrequencies $\omega_{\alpha}$, would be brought into resonance when $\Oo$ falls in the interval $[ \frac{1-\epsilon}{2}\omega_{\alpha},\frac{1+\epsilon}{2}\omega_{\alpha} ]$ \citep{Lai:1993di}.
In our numerical results, we find that the definition
\begin{align}\label{eq:resnbh}
	\epsilon=10\sqrt{  \frac{2\pi }{\Omega_{\text{orb}}}  \frac{|\dot{a}|}{a}     } 
\end{align} 
is adequate for determining the onset and the end of resonance (see Fig.~\ref{fig:ampres}), which in turn yields the resonance duration $t_{\text{res}} \approx \epsilon \text{Re}(\omega_{\alpha})$.
On the other hand, the tidal field shifts the eigenfrequencies by (see also Sec.~\ref{sec.V.B})
\begin{align}
	\delta\omega^{T}_{\alpha}=\frac{Q_{n2} }{2\omega_{\alpha}a^{3}}\Mc,
	\label{eq:modtid}
\end{align}
where $\omega_{\alpha}$ is the unperturbed eigenfrequency.
This indicates that the true eigenfrequencies --- which include a shift due to the tidal field --- must be solved for simultaneously with the orbital evolution equations, since Eq.~\eqref{eq:modtid} depends on the (time-dependent) orbital separation $a$.
In fact, any perturbing force will shift the pulsation spectrum because the kinetic and potential operators $\mathcal{V}$ and $\mathcal{T}$ defined in equation \eqref{eq:operat} are adjusted accordingly. In Sec.~\ref{sec.V}, we show how the Coriolis and Lorentz forces associated with a rotating and magnetized star may influence the spectrum.

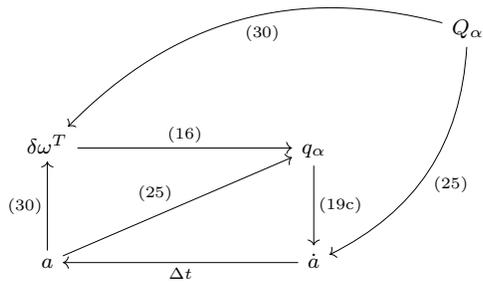
\begin{figure} 
	\centering
	\begin{tikzcd}[row sep=huge, column sep=huge]
		& & & Q_{\alpha} \arrow[dlll, bend right, "\eqref{eq:modtid}"] \arrow[ddl, bend left, "\eqref{eq:Htid}"]  \\
		\delta\omega^{T} \arrow[rr, "\eqref{eq:osc}"] & & q_{\alpha} \arrow[d, "\eqref{eq:eomorb}"] \\
		a \arrow[u, "\eqref{eq:modtid}"] \arrow[urr, "\eqref{eq:Htid}"]& &\dot{a} \arrow[ll, "\Delta t"]&
	\end{tikzcd}
	\caption{The numerical scheme used in this paper to evolve the modes of the primary.
	The only time-independent quantities are the tidal overlap integrals $Q_{\alpha}$ on the upper right, while the rest are iteratively solved for. 
	Starting from the separation $a(t)$, the strength of tidal field by the companion of a specific binary is decided and gives rise to certain shift in eigenfrequencies of QNMs [Eq.~\eqref{eq:modtid}], which depends also on the tidal coupling strength of each QNM $Q_{\alpha}$.
	Consequently, total eigenfrequencies $\omega_{\alpha}+\delta\omega^{T}_{\alpha}$ fixes the Hamiltonian of QNMs [Eq.~\eqref{eq:osc}], which is solved to update mode amplitudes.
	Next, the change rate of the separation $\dot{a}$ is influenced by the amplitude of excited pulsations and the tidal coupling strength $Q_{\alpha}$, and infers the separation at the next moment $a(t+\Delta t)$ for the time step $\Delta t$. 
	Then the cycle runs again until $a\lesssim3\Rs$ \citep{Ho99,pap1}.
	Each arrow stands for a deduction via the relation labeled beside it. }
	\label{diag:tid}
\end{figure}

\begin{figure} 
	\centering
	\includegraphics[scale=0.4]{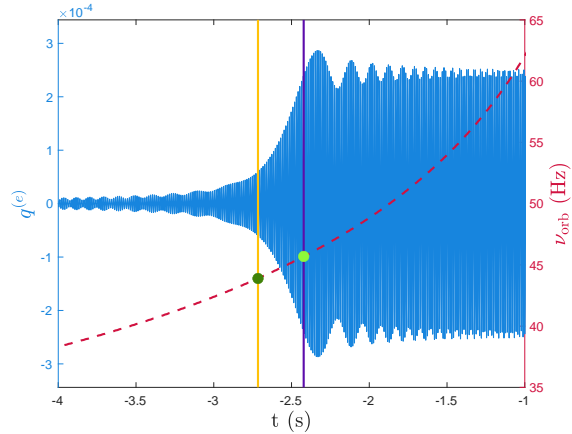}  
	\caption{The amplitude $q^{(e)}$ (blue; left y-axis) of the $l=m=2$, $g_1-$mode, whose unperturbed frequency is $89.25$ Hz, and the orbital frequency (red; right y-axis) as functions of time.
	The horizontal axis records the time prior to an equal mass NS-NS coalescence, which is achieved once the separation decays to $a\lesssim 3 \Rs$ \citep{Ho99,pap1}.
	The yellow and the purple lines mark the beginning and the end of the resonance respectively; the corresponding orbital frequencies are $43.89$ Hz and $45.69$ Hz, marked by solid green points.
	We have taken an equal-mass binary with the SLy EOS with $\Ms=1.27M_\odot=\Mc$ and $\Rs=11.78$ km.
	The radial displacement $\xi^r$ of the first five $g$-modes are shown in Fig.~\ref{fig:gmodes}.}
\label{fig:ampres}
\end{figure}

The numerical scheme for evolving the modes of the primary is summarised in Fig.~\ref{diag:tid}.
We begin by evolving the binary with an initial separation of 
\begin{align}
	a(0) = 2\sqrt[3]{\frac{\Ms+\Mc}{ \big(\text{Re}[\omega_{\alpha}]\big)^{2}}}.
	\label{eq:inisep}
\end{align}
The initial (non-resonant) mode amplitude is assumed to be zero.
In Fig.~\ref{fig:ampres} we show the mode amplitude for the $l=m=2$, $g_1-$mode for a star with SLy EOS as a member of an equal-mass $(q=1)$ binary as a function of time.
The resonance starts at orbital frequency $\nu_{\text{orb}}=\Omega_{\text{orb}}/2\pi=43.89$ Hz (green point) and ends at $45.69$ Hz (light green point) with duration of $0.29 \text{ s}$.
The mode oscillates with amplitude $q \approx2.4\times10^{-4}$ after resonance, though these oscillations decay exponentially as merger is reached.

We performed simulations for equal-mass binaries assuming the EOS APR4, SLy, and WFF1-3.
We find that the maximal amplitudes of $g_1-$modes of the primary obey the following approximate relations
\begin{align}\label{eq:maxamp}
	q_{\alpha,\text{max}} (\omega_{\alpha}\Ms)^{5/6} \simeq (0.1092\pm 0.0208) Q_{12}, 
\end{align}
where the error is given by  1$\sigma$ confidence level of least squares fitting.
The analytic equation of the maximum mode amplitude under the stationary phase approximation suggests a slope of $\pi/32 \simeq 0.0982$ [cf.~Eq.~(6.3) in \cite{Lai:1993di}], which agrees our result to within the stated confidence level.

\begin{figure}
	\centering
	\hspace*{-.5cm}\includegraphics[width=0.57\textwidth]{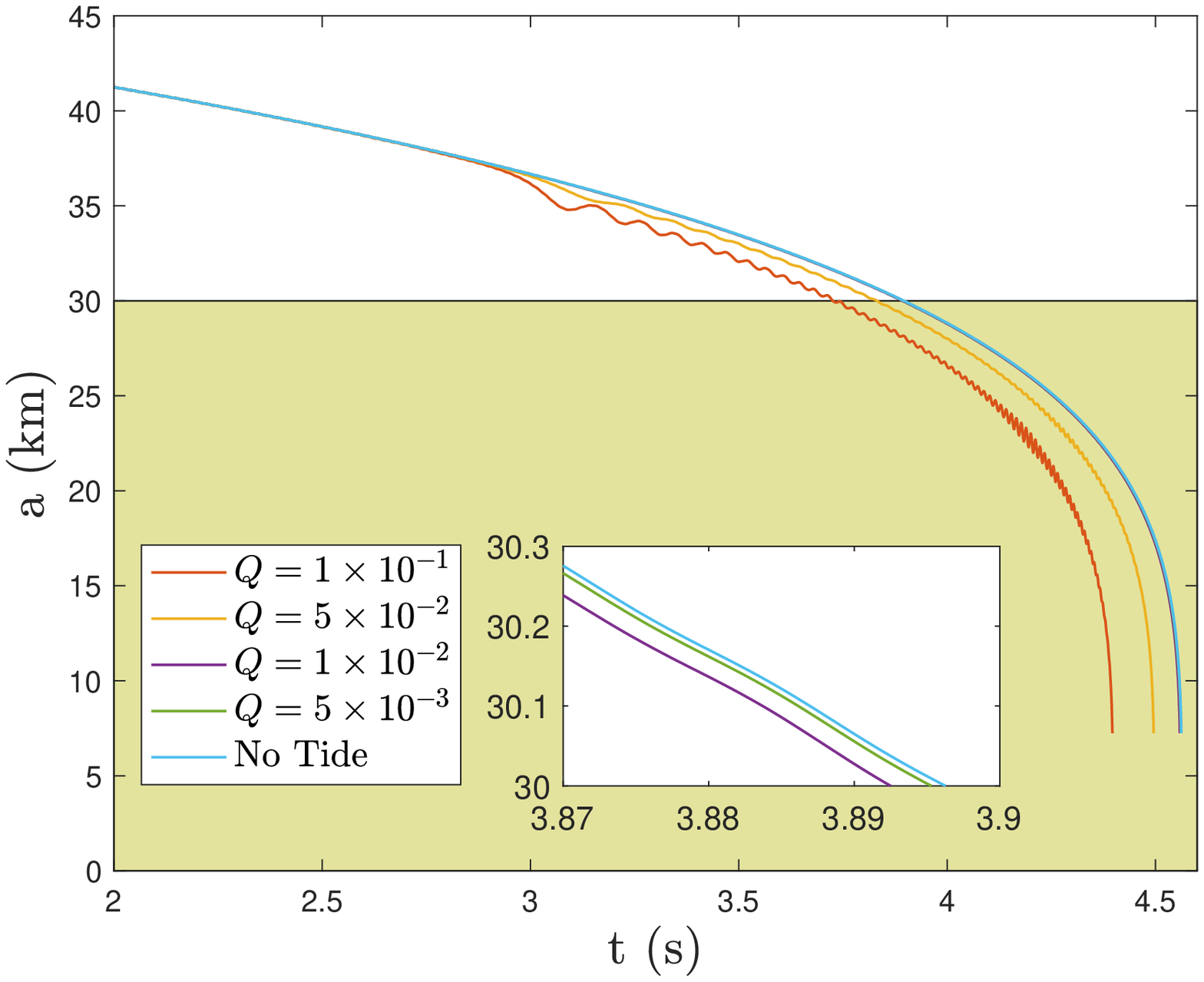}
	\caption{ Binary separation with various tidal overlaps as a function of time. 
	The maximal disagreement between $Q=0.01$ and the case without tidal effect is $\lesssim 0.01$ s and is smaller for smaller tidal overlaps.
	Shaded area, where $a\lesssim 3\Rs$ \citep{Ho99,pap1}, marks the stage after the merger, which thus is not the virtual evolution.
	We have taken $\Ms=1.4M_{\odot}=\Mc$ and $\Rs=10$ km, hence the collapse happens when the separation is around 30 km. The frequency of the resonant mode is 100 Hz. This Figure does not use any particular EOS.
	}
\label{fig:orbit}
\end{figure}

Tidal effects accelerate the merger because orbital energy leaks into the QNMs \citep{Kokkotas:1995xe,Vick:2019cun}, especially the $f$-modes, whose coupling strengths are typically a few tenths. For modes with coupling strengths $Q \lesssim 0.01$, the effects on the orbital evolution are negligible.
In Fig.~\ref{fig:orbit}, we present the separation of an equal-mass binary with $\Ms=1.4M_{\odot}=\Mc$ and $\Rs=10$ km  for four different strengths of tidal overlap $Q$ as functions of time, together with an evolution on which the tidal effects are absent.
As such, one can observe that for $Q\lesssim 0.01$ the tidal effect on the evolution is quite small. 
Given that the typical coupling strengths of a $g-$mode are both much smaller than 0.01, the effect in this case of the $g-$mode resonances on the orbital evolution are insignificant relative to measurement uncertainties in the timing of GWs and gamma ray-bursts.

\section{Magnetohydrodynamics}
\label{sec.IV}

Having explored the effect of the $g-$modes on the orbital evolution, we now turn to the influence of magnetic fields. 
In this work, we treat stars as perfect conductors over which relativistic magnetic fields are constructed. 
Perturbed magnetic fields generate Lorentz forces according to the Faraday induction equation (Sec.~\ref{sec.IV.B}), which tunes the eigenfrequencies of QNMs through back-reaction (Sec.~\ref{sec.V.A}). 
In the event that a crustal yielding occurs on a magnetized star,  the release of fracture energy may generate flares, such as precursors of short gamma-ray bursts. In general, perturbed magnetic fields would induce electric fields, accelerating charged particle and thermalizing the electromagnetic emission; however, if the magnetic field is strong enough [$\gg 10^{13}$ G, i.e., magnetar-level \citep{Duncan1992}], the energy propagates along field lines as Alfv{\'e}n waves \citep{Tsang:2011ad}. The non-thermal properties of precursors thus support the possibility that at least one NS is a highly magnetized star in those events \citep{pap1}. In any case, timing and the spectral properties of precursor will be examined in Paper II as an application of the formalism developed here.

\subsection{Magnetic field structure}\label{sec.IV.A}

The Viral theorem \citep{Chandrasekhar53} sets an upper limit to the magnetic field strength for NSs of the order $\sim 10^{18}$ G \citep{Lai01,Lattimer07,Reisenegger:2008yk}.
Even for most magnetars, the (surface dipole) magnetic field strength is much smaller than this extreme, implying that the gravitational binding energy exceeds the magnetic energy by several orders of magnitude \citep{Sotani:2006at,Akgun:2013}. One may therefore treat the magnetic field as a perturbation over a spherically symmetric background profile \eqref{eq:statsphsym}, in the style of \cite{ciolf,mast11,mast15}. 

We introduce the electromagnetic 4-potential $A_{\mu}$, which defines the Faraday tensor
\begin{equation}
F_{\mu \nu} = \nabla_{\nu} A_{\mu} - \nabla_{\mu} A_{\nu},
\end{equation}
where each $A_{\mu}$ is a function of $r$ and $\theta$ only. Maxwell's equations for the electromagnetic field are
\begin{equation}\label{eq:Maxwell}
F^{\mu \nu}_{;\nu} = 4 \pi J^{\mu} , \hspace{0.4cm} \nabla_{[ \alpha} F_{\beta \gamma ]} = 0,
\end{equation}
for 4-current $J^{\mu}$ [effectively defined by the first of equations.~\eqref{eq:Maxwell}]. The Lorentz force is then given by $F^{\mu}_{L} = F^{\mu \nu} J_{\nu}$. The ideal MHD condition of vanishing electric field, defined by
\begin{equation} \label{eq:elecfield}
E_{\mu} = F_{\mu \nu} u^{\nu},
\end{equation}
for a static and non-rotating star (i.e.~$u^{\mu} = e^{-\Phi} \partial_{t}$), returns the condition $A_{t} = 0$. 
We have residual gauge freedom, which allows us to pick $A_{\theta} = 0$ \citep{glamp}. 
Setting $A_{r} = \Bc e^{\lambda - \Phi} \Sigma$ and $A_{\phi} = \Bc \psi$, it can be shown that Maxwell's equations are solved exactly for \citep{ciolf}
\begin{equation}
\Sigma(r,\theta) = \int d \theta \zeta(\psi) \frac {\psi(r,\theta)} {\sin \theta},
\end{equation}
for some $\zeta$, which is an arbitrary function of the stream function $\psi$ and effictively defines the azimuthal (toroidal) component $B_{\phi}$, effectively generalising the Chandrasekhar (Helmholtz) decomposition in flat space \citep{Chandrasekhar56,mast11}.
Here $\Bc$ sets the characteristic field strength.

The magnetic 4-field has covariant components
\begin{equation} \label{eq:covmag}
B_{\mu} = \frac{1} {2} \epsilon_{\mu\nu\sigma\eta} u^{\nu} F^{\sigma \eta},
\end{equation}
where $\boldsymbol{\epsilon}$ denotes the Levi-Civita symbol.
Using the above expression for $A_{\mu}$, the contravariant components $B^{\mu}$ can be readily evaluated, and we find 
\begin{equation} \label{eq:magf}
B^{\mu} = \Bc \left( 0, \frac {e^{-\lambda}} {r^2 \sin\theta} \frac{\partial\psi}{\partial\theta}, 
- \frac {e^{-\lambda}} {r^2 \sin\theta}  \frac{\partial\psi}{\partial r}, 
- \frac {\zeta(\psi) \psi e^{-\Phi}} {r^2 \sin^2\theta}  \right). 
\end{equation}
The function $\psi$ can now be expanded as a sum of multipoles. For simplicity, we take a dipole field with polynomial radial component [which generalises the Newtonian description in \cite{mast11}], i.e., 
\begin{equation}
\psi(r,\theta) = f(r) \sin^2\theta,
\end{equation}
with $f(r) = a_{1} r^2 + a_{2} r^4 + a_{3} r^6$, where the $a_{i}$ are to be constrained by appropriate boundary conditions.  
In particular, we impose that (i) the field matches to a force-free dipole outside of the star ($r>\Rs$),  and (ii) there are no surface-currents ($J^{\mu}|_{r=R_\star}=0$). This leads to \emph{four} constraints,  which are that the 3-components \eqref{eq:magf} are continuous at the boundary $\partial V$  of the star, and that the 4-current vanishes there (which only has one non-trivial component, $J_{\phi}$, for an axisymmetric field). One of these is trivially satisfied by demanding that $\zeta$ vanishes on the surface, which we achieve, as in \cite{mast11}, by setting
\begin{equation}
\zeta(\psi) \psi = - \left[ \frac{E^{p} \left( 1 - \Lambda \right)} {E^{t} \Lambda} \right]^{1/2}
 \frac{\left( \psi - \psi_{c} \right)^2}{\Rs^3},
 \label{eq:torratio}
\end{equation}
when $\psi \geq \psi_{c}$, and $\zeta$ is zero otherwise.
Here $\psi_{c}$ is the critical value of the streamfunction, defined as the value of the last poloidal field line that closes within the star, thus the toroidal component is confined to the region of closed poloidal field lines.
The quantity $\Lambda$ measures the ratio of poloidal and toroidal magnetic energies; typically $\Lambda \ll 1$ for a stable configuration \citep{Braithwaite09,Akgun:2013}.
For the above choices, we find
\begin{equation}
\psi_{c} = - \frac {3 \Rs^3 \sin^2\theta} {8 \Ms^3} \left[ 2 \Ms \left( \Ms + \Rs \right) + \Rs^2 \log \left( 1 - \frac{2 \Ms} {\Rs} \right) \right].
\end{equation}
The energy stored in the \emph{internal} magnetic field of the static equilibrium is [see, e.g., Eq.~(41) in \cite{ciolf}]
\begin{align}
	E = 2\mathlarger{\int}_{\text{primary}} \sqrt{-g}d^{3}x u_{\mu}u_{\nu}T^{\mu\nu}
	=\frac{1}{4\pi}\mathlarger{\int}_{\text{primary}} \sqrt{-g}d^{3}x  B^{2},
\end{align}
where $T^{\mu\nu}$ is the magnetic stress-energy tensor
\begin{align}
	T^{\mu\nu}=\frac{B^{2}}{4\pi}\bigg( u^{\mu}u^{\nu}+\frac{1}{2}g^{\mu\nu} \bigg)-\frac{B^{\mu}B^{\nu}}{4\pi},
\end{align}
with $B^{2} = B^{\mu}B_{\mu}$. 
For the dipolar field \eqref{eq:magf} considered here, the poloidal and toroidal energies are
\begin{align}
	E^{p} = \frac{\Bc^2}{4\pi}\int_{\text{primary}}\sqrt{-g}d^{3}x \bigg[ 
	\bigg( \frac{\partial_{\theta}\psi}{r^{2}\sin\theta} \bigg)^{2} +
	\bigg( \frac{e^{-\lambda}\partial_{r}\psi}{r\sin\theta} \bigg)^{2}\bigg],
\end{align}
and
\begin{align}
	E^{t} =\frac{\Bc^2}{8\pi} \int_{\psi\ge\psi_{c}} \sqrt{-g}d^{3}x 
	\bigg( \frac{\zeta(\psi)\psi e^{-\Phi}}{r\sin\theta} \bigg)^{2},
\end{align}
 respectively.

The force-free dipole outside of the star is found by setting $F^{\mu}_{L} = 0$ for $r > \Rs$. 
This leads to \citep{Wasserman:1983}
\begin{equation}
\psi_{\text{ext}} = -\frac {3 \Rs^3 \sin^2 \theta} {8 \Ms^3} \bigg[ 2 \Ms (r + \Ms)+r^2\log\bigg(1-\frac{2\Ms}{r}\bigg) \bigg],
\end{equation}
where we note that, outside of the star, the geometry is Schwarzschild, i.e. 
\begin{equation}
\Phi(r>\Rs) = \frac{1}{2} \log \bigg( 1- \frac{2M}{r}\bigg) \,\, \text{and} \,\, \lambda(r>\Rs) = -\Phi.
\end{equation} 
It is not hard to prove (use L`Hopital's rule) that, in the limit $\Ms \rightarrow 0$, $\psi_{\text{ext}}$ reduces to the standard force-free dipole of Newtonian theory, $\psi \sim \sin^2\theta / r$. 
Finally, imposing the conditions (i) and (ii) discussed above leads to
\begin{subequations}
\begin{align}
a_{1} =& - \frac {3 \Rs^3} {8 \Ms^3} \Big[ \log \left( 1 - \frac{ 2 \Ms} {\Rs} \right) \nonumber\\
&+ \frac {\Ms \left( 24 \Ms^3 - 9 \Ms^2 \Rs - 6 \Ms \Rs^2 + 2 \Rs^3 \right)} {\Rs^2 \left( \Rs - 2 \Ms \right)^2} \Big], \label{eq:a1}\\
a_{2} =& \frac {3 \left( 12 \Ms - 7 \Rs \right)} {4 \Rs \left( \Rs - 2 \Ms \right)^{2}}, \label{eq:a2}
\end{align}
and lastly
\begin{align}
a_{3} = \frac {3 \left( 5 \Rs - 8 \Ms \right)} {8 \Rs^3 \left( \Rs - 2 \Ms \right)^{2}}. \hspace{3.3cm} \label{eq:a3}
\end{align}
\end{subequations}
The above therefore completely defines the general-relativistic generalisation of the \cite{mast11} mixed poloidal-toroidal field.

\begin{figure}
	\centering
	\hspace*{-0.3cm}\includegraphics[width=0.5\textwidth]{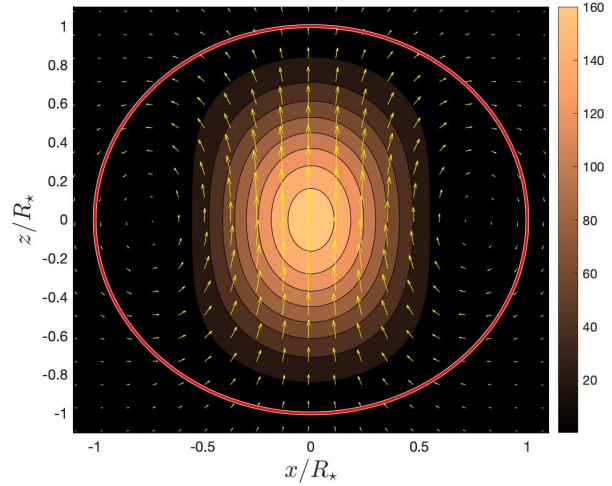}
	\caption{ Field lines for the background magnetic field $\mathbf{B}$ with $\Lambda = 1$. The red rigid
	line marks the surface of the equilibrium star; the scale is given in unit of $\Bc$ for which brighter shades indicate stronger field strength.
	}
\label{fig:magprofile}
\end{figure}

The magnetic field introduces a frequency shift in the spectrum of the star depending on the values $\Bc$ and $\Lambda$, defining the characteristic poloidal and toroidal strengths. To better understand the magnetic field, we transform the contravariant components of 4-field $B^{\mu}$ into the \emph{Newtonian-like} components, denoted by a overhead tilde, through [cf.~Eq.~(4.8.5) in \cite{Weinberg72}]
\begin{align}
	\tilde{B}_{a} = \sqrt{g_{aa}}B^{a},
\end{align}
and we show the cross section in Fig.~\ref{fig:magprofile}, which is the whole picture of the magnetic field 
if the field is purely poloidal. 

\subsection{Perturbed Lorentz force and Faraday equations}\label{sec.IV.B}

We now study the backreaction onto the magnetic fields induced by the modes, which results in the frequency modulation \ref{sec.V}. In this section, we ignore the spacetime variation of QNMs \emph{from magnetic fields} in the perturbation equations ($\delta g^{\text{mag}}_{\mu\nu} = 0$). Nonetheless we note that the difference between our approximation and the Cowling one is that we take the first order spacetime perturbations into account, viz.~spacetime perturbations are involved in determining the mode frequency of the stars, and only the higher-order `magnetic-backreaction' effects are disposed of. In this sense, our results are expected to be more accurate than those which adopt the Cowling approximation.

Following the derivation in \cite{Sotani:2006at} [see also \cite{Papadopoulos:1982}], the projection of the equation of motion onto the hypersurface orthogonal to $u^{\mu}$,
\begin{align}
	h^{\mu}_{\eta}\nabla_{\nu}T^{\eta\nu}=0,
\end{align}
gives
\begin{align}\label{EOM}
(\rho+p+\frac{B^{2}}{4\pi}) u^{\nu} \nabla_{\nu}u^{\mu} 
=-h^{\mu\nu}  \nabla_{\nu}\bigg(p+\frac{B^{2}}{8\pi}\bigg)
+h^{\mu}_{\eta}  \nabla_{\nu}\bigg( \frac{B^{\eta}B^{\nu}}{4\pi} \bigg),
\end{align}
where the projection operator $h^{\mu\nu} = g^{\mu\nu}+u^{\mu}u^{\nu}$.
The Lagrangian 4-displacement $\xi^{\mu}$ is related to the perturbed velocity through the Lie derivative [see, e.g., Eq.~(34) of \cite{fried}]
\begin{equation}\label{Lag2vel}
\delta u^{\mu} = h^{\mu}_{\nu} \mathcal{L}_{\boldsymbol{u}} \xi^{\nu}.
\end{equation}
In the simple case of a static fluid ($u^{\mu}=e^{-\Phi}\partial_t$), we find $\delta u^{i} = i \omega_{\alpha} e^{- \Phi} \xi^{i}$ for a certain QNM.
After linearizing the equation of motion \eqref{EOM} and utilizing relation \eqref{Lag2vel}, the perturbing Lorentz force reads
\begin{align}\label{eq:Lorentzpert}
\delta F^{\mu}_{B} =&	\frac{i\omega_{\alpha} e^{-\Phi}}{4\pi}\bigg[ B^{2}\bigg( \nabla_{t}(\xi^{\mu}e^{-\Phi}) 
+\xi^{\nu}\nabla_{\nu}u^{\mu}  \bigg) \nonumber\\
&+\frac{u^{\mu}\xi^{\nu}+u^{\nu}\xi^{\mu}}{2} \nabla_{\nu}(B^{2})
-(u^{\mu}\xi_{\eta} e^{2\Phi} +\xi^{\mu}u_{\eta})\nabla_{\nu}(B^{\eta}B^{\nu}) \bigg] \nonumber\\
&-\frac{1}{4\pi}h^{\mu}_{\eta}\nabla_{\nu} (B^{\eta}\delta B^{\nu}  +\delta B^{\eta}B^{\nu} )
+\frac{1}{4\pi}h^{\mu\nu}\nabla_{\nu} (B_{\eta}\delta B^{\eta}) \nonumber\\
&+\frac{1}{2\pi}B_{\eta}\delta B^{\eta}u^{\nu}\nabla_{\nu}u^{\mu},
\end{align}
while the perturbed magnetic field $\delta B^{\mu}$ can be determined by solving the linearized induction equation \citep{Sotani:2006at}
\begin{align}
\nabla_{t}\delta B^{\mu}=& i\omega_{\alpha} \bigg[ 
-\xi^{\nu}\partial_{\nu}B^{\mu}
-\xi^{\nu}\Gamma^{\mu}_{\nu\eta}B^{\eta}
-u^{\mu}\xi_{\nu}B^{\eta}\nabla_{\eta}u^{\nu} \nonumber\\
&+e^{\Phi}B^{\nu}\nabla_{\nu}(e^{-\Phi}\xi^{\mu})
-e^{\Phi}B^{\mu}\partial_{\nu}(e^{-\Phi}\xi^{\nu})
-B^{\mu}\Gamma^{\nu}_{\nu\eta}\xi^{\eta} \nonumber\\
&+u^{\mu}B^{\nu}i\omega_{\alpha} e^{-\Phi}\xi_{\nu}
+u^{\mu}B^{\nu}\xi^{\eta}   (\partial_{\eta}u_{\nu}+\Gamma^{t}_{\nu\eta}e^{\Phi}) \nonumber\\
&-u^{\mu}B^{\nu}\Gamma^{\eta}_{\nu t}\xi_{\eta}
+\xi^{\mu}B^{r}\Phi'
\bigg] +\delta B^{r}u^{\mu}\Phi'e^{\Phi}.
\end{align}
For the magnetic field given by Eq.~\eqref{eq:magf}, the induction equation gives
\begin{align}\label{Faraday}
\frac{\partial}{\partial t}\delta B^{\mu}=
&i\omega_{\alpha} \bigg[ 
-\xi^{r}\partial_{r}B^{\mu}-\xi^{\theta}\partial_{\theta}B^{\mu}
+B^{\nu}\partial_{\nu}\xi^{\mu}+B^{\mu}\Phi'\xi^{r} \nonumber\\
&-B^{\mu}\partial_{\nu}\xi^{\nu}-B^{\mu}\Gamma^{\nu}_{\nu\eta}\xi^{\eta} +u^{\mu}B^{\nu}i\omega_{\alpha} e^{-\Phi}\xi_{\nu} \bigg] \nonumber\\
&+\delta B^{r}u^{\mu}e^{\Phi}\Phi'-\Gamma^{\mu}_{t\nu}\delta B^{\nu}.
\end{align}
Since some terms only appear in the temporal component of the first-derivative of $\delta B^{\mu}$, one can make the equations more concise by separating the temporal component from the spatial, viz.
\begin{subequations}
\begin{align}
\frac{\partial}{\partial t}\delta B^{t}=&-\omega_{\alpha}^{2}  e^{-2\Phi} B^{\nu}\xi_{\nu}, \\
\frac{\partial}{\partial t}\delta B^{i}=& -i\omega_{\alpha} \bigg[ 
\bigg(\xi^{r}\partial_{r} + \xi^{\theta}\partial_{\theta}-\Phi'\xi^{r}+\partial_{\nu}\xi^{\nu}  \nonumber\\
&+ \xi^{\nu}\partial_{\nu}\ln\sqrt{|g|}\bigg)B^{i}
-B^{j}\partial_{j}\xi^{i}\bigg].
\end{align}
\end{subequations}
Accordingly, one can integrate the above equations to find
\begin{subequations}\label{perturbB}
	\begin{align}
	\delta B^{t}= i\omega_{\alpha}  e^{-2\Phi}B^{a}\xi_{a}, \hspace{4.2cm}
	\end{align}
	\begin{align}
	\delta B^{i}=& -\bigg[ 
	\bigg(\xi^{r}\partial_{r} + \xi^{\theta}\partial_{\theta}-\Phi'\xi^{r}+ \partial_{a}\xi^{a}  \nonumber\\
	&+\xi^{r}\bigg(\Phi'+\lambda'+\frac{2}{r}\bigg)+\frac{\xi^{\theta}}{\tan\theta}	\bigg)B^{i}
	-B^{j}\partial_{j}\xi^{i} \bigg].
	\end{align}
\end{subequations}
Expressions \eqref{perturbB} will be used to define the perturbing Lorentz force in Sec.~\ref{sec.V.A}.

Though it is not considered in this work, we would like to point out the possibility that NSs as a member of a binary may be cold enough that they contain superconducting matters, which influences the magnetic properties of the star \citep{Lander13} and any resulting GWs \citep{suv21}. For example, the force induced from the field perturbation has different influences to the Lorentz force, and the induction equation is also altered due to the altered nature of Ohmic and ambipolar dissipation \citep{Graber15}. On top of that, superfluidity increases the frequencies of $g$-modes, e.g., \cite{Yu17} found that the frequency of the $n=1$, $g_1$-mode is of the order of a few hundred Hz (even up to 700 Hz) for a particular EOS with $M=1.4M_{\odot}$ (see their Fig.~4). Adding that the overlap integral is found to be of the same order as the case where the superfluid is absent (see their Fig.~5), we expect a smaller amplitude for the $g_1$-mode due to its shorter resonance timescale. Although the inclusion of superfluidity brings higher-order modes into play, those overtones typically have a much weaker overlap integral (with respect to the normal fluid case). It may thus not be very plausible that these overtones can account for tidally-driven crustal failure.


\section{Mode frequency modulations}
\label{sec.V}
The introduction of a perturbing force $\delta F^{\mu}$ into the Euler equations \eqref{eq:euler} leads to a modulation $\delta \omega$ in mode frequencies, while eigenfunctions are left unchanged to leading order \citep{Unno:1979,Bi03,pap1}.
The restriction of the Euler equation \eqref{eq:euler} of the unperturbed equilibrium to the hypersurface orthogonal to $u^{\mu}$, i.e.~$h_{\mu\nu}\nabla_{\eta}T^{\eta\nu}=0$, gives
\begin{align}
	(\Po) u^{\nu}\nabla_{\nu}u^{\mu}=-h^{\mu\nu}\nabla_{\nu}p,
\end{align}
from which and Eq.~\eqref{Lag2vel}, one can derive the linearized equation
\begin{align}\label{eq:eigenprob}
(\Po)e^{-2\Phi}\omega^{2}\xi^{\mu}=&
(\delta\rho+\delta p) u^{\nu}\nabla_{\nu}u^{\mu} +h^{\mu\nu} \nabla_{\nu}\delta p \nonumber\\
&+i\omega e^{-\Phi}\bigg[ (\Po) \nabla_{\nu}u^{\mu} +u^{\mu}\nabla_{\nu}p \bigg]\xi^{\nu}  \nonumber\\
&+i\omega e^{-\Phi}\xi^{\mu}u^{\nu}\nabla_{\nu}p.
\end{align}
The left hand side and the right hand side give, respectively, the kinetic operator $\mathcal{T}$ and the potential operator $\mathcal{V}$ that are defined in Eq.~\eqref{eq:operat}. This equation, with appropriate boundary conditions, forms an eigenvalue problem for $\omega_{\alpha}^{2}$.

For an mode with unperturbed frequency $\omega_{\alpha}$, the inclusion of a perturbing force $\delta F^{\mu}$ on the right hand side of \eqref{eq:eigenprob}, its eigenvalues would be amended accordingly by $\delta\omega_{\alpha}$.
Substituting $\omega=\omega_{\alpha}+\delta\omega_{\alpha}$ and focusing on the leading order perturbation terms, \eqref{eq:eigenprob} gives
\begin{align}
	2(\Po)e^{-2\Phi}\omega_{\alpha}\xi^{\mu} \delta\omega_{\alpha}=\delta F^{\mu},
\end{align}
from which the frequency shift,
\begin{align}\label{eq:freqpert}
\delta\omega_{\alpha}=  \frac{1}{2\omega_{\alpha}}
\frac{\mathlarger{\int_{\text{primary}}\delta F_{\mu}\bar{\xi^{\mu}} \sqrt{-g}d^{3}x}}
{\mathlarger{\int_{\text{primary}}(\rho+p)e^{-2\Phi}\xi^{\mu}\bar{\xi_{\mu}}\sqrt{-g}d^{3}x}},
\end{align}
can be obtained. A similar derivation in the Newtonian case can be found in \cite{Unno:1979}. Equation \eqref{eq:freqpert} is numerically evaluated for some particular choices of $\delta F^{\mu}$.

\subsection{Magnetic field} \label{sec.V.A}
As $\delta \boldsymbol{F}$ is given by the Lorentz force, \eqref{eq:Lorentzpert} and \eqref{eq:freqpert} yield the expression of the correction in the frequency for a general magnetic field, which, after substituting the magnetic field as defined in Eq.~\eqref{eq:magf} and adopting the normalization \eqref{eq:normalization}, becomes
\begin{align}
\delta\omega^B_{\alpha}
=&\frac{(\Ms\Rs^{2})^{-1}}{8\pi\omega_{\alpha}}\int_{\text{primary}} \sqrt{-g} d^{3}x \bigg[
-\omega_{\alpha}^{2}B^{2}\xi^{\mu}\bar{\xi}_{\mu}e^{-2\Phi} \nonumber\\
&+2B_{\mu}\delta B^{\mu}\bar{\xi}^{r}\Phi' 
-\bar{\xi}_{\mu}\nabla_{\nu}\bigg(B^{\mu}\delta B^{\nu}+B^{\nu}\delta B^{\mu}\bigg) \nonumber\\
&+\bar{\xi}^{\nu}\nabla_{\nu}(B_{\mu}\delta B^{\mu})
\bigg].
\label{eq:modmag}
\end{align}

\begin{figure} 
	\centering
	\includegraphics[scale=0.45]{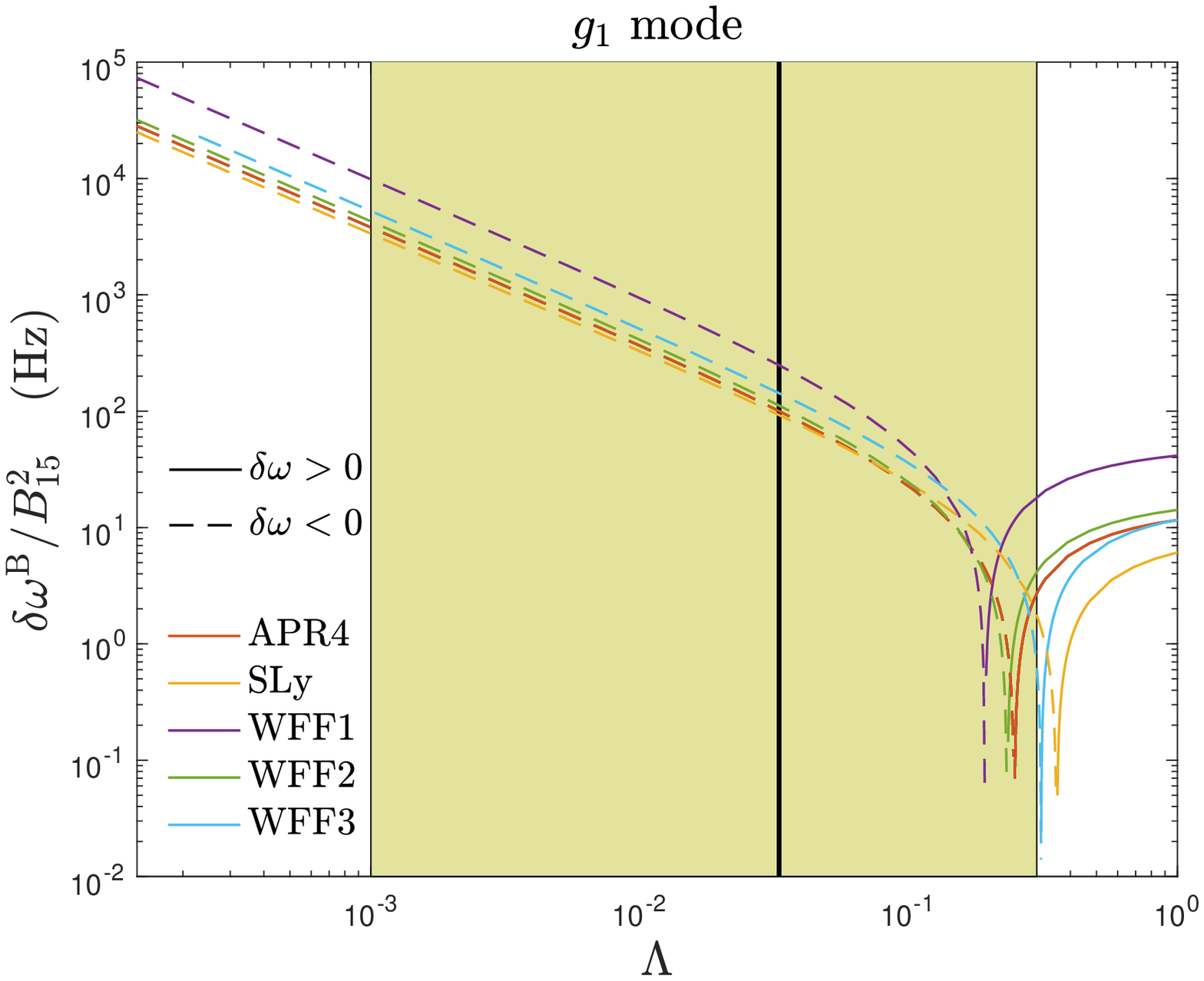} 
	\includegraphics[scale=0.45]{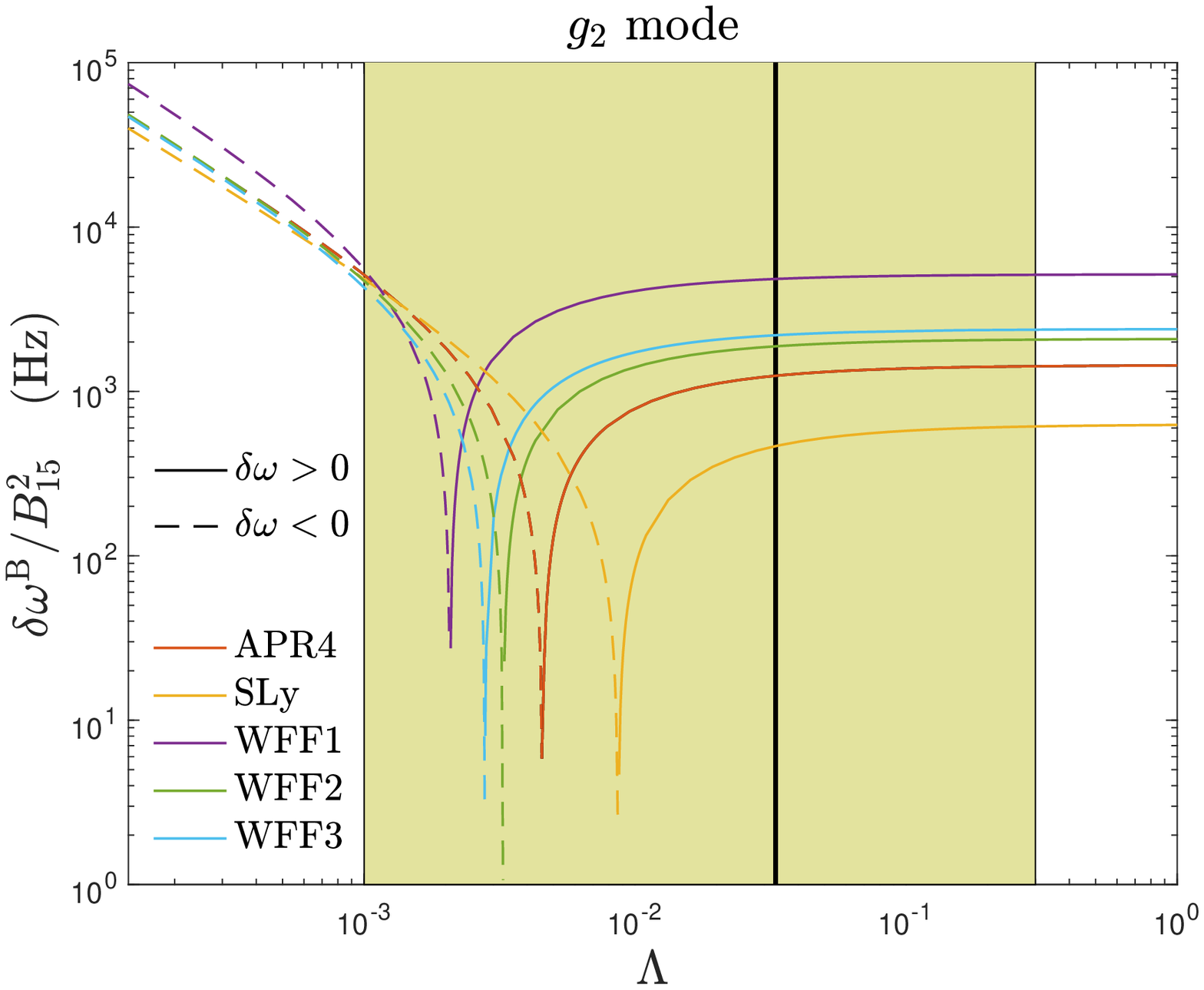} 
	\caption{Eigenfrequency shifts $\delta\omega^{B}$ for the $l=2$, $g_{1}-$ (top panel) and $g_{2}-$modes (bottom panel) due to the magnetic field as functions of the poloidal-to-toroidal field-strength ratio $\Lambda$. The shaded areas represent the range of $\Lambda$ for which the magnetic field is stable \citep{Akgun:2013}. The black solid lines mark the ratio $\Lambda=0.033$, which gives the maximal toroidal field strength for $\Bc=10^{15}\text{ G}$ inside the star [Eq.~\eqref{eq:torlimit}].
		We have used $\delta=0.005$ and $\rho_{c}=9\times10^{14}$ g/cm$^{3}$ for EOS APR, SLy, and WFF1-3, whose masses are 1.21$M_{\odot}$, 1.27$M_{\odot}$, 0.86$M_{\odot}$, 1.14$M_{\odot}$, and 1.04$M_{\odot}$, respectively.
	}
	\label{fig:p2t}
\end{figure}

\begin{figure} 
	\centering
	\includegraphics[scale=0.43]{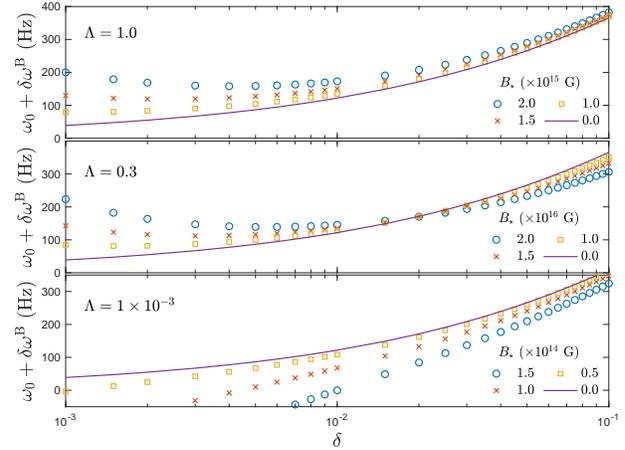}  
	\caption{ $g_1-$mode frequencies: the purple continuous line is for unperturbed star, while the rest for magnetized ones with different strengths of the magnetic field $\Bc$ as a function of $\delta$.
	We set $\Lambda=1.0$ (top panel), $\Lambda=0.3$ (middle panel), and $\Lambda=10^{-3}$ (bottom panel).
	The background star is constructed by EOS APR4 and has the central density of $8\times10^{14}$ g/cm$^{3}$.
	}
	\label{fig:bfreqAPR}
\end{figure}

In Fig.~\ref{fig:p2t}, we plot the mode frequency shifts for the $l=2$, $g_1-$ and $g_2-$modes ($n=2$) with some fixed stellar parameters and EOS as functions of the poloidal-to-toroidal strength $\Lambda$ (top panel). The range of $\Lambda$ is chosen broadly compared to the ratio for a stable magnetic field configuration (shaded area), which is $10^{-3} \lesssim \Lambda \lesssim 0.3$ \citep{Akgun:2013,Herbrik17}. The stability examined by the energy variation method gives the constraint
\begin{align}\label{eq:torlimit}
	\tilde{B}^{\phi} \lesssim 10^{17} \sqrt{\Bn} \bigg( \frac{\delta}{0.01} \bigg) \text{ G},
\end{align}
on the toroidal strength, which implies $\Lambda\gtrsim0.033$ (black line) for a magnetar-level surface field strength $B_{\star} \sim 10^{15} \text{ G}$.
This constraint on the strength of toroidal component becomes loose for larger $\delta$ \citep{Akgun:2013,Herbrik17}.
On the other hand, we find that the Virial limit on the field strength of $\sim10^{18}$ G inside the star  \citep{Lai01,Lattimer07,Reisenegger:2008yk} corresponds to $\Lambda \gtrsim 5 \times 10^{-4}$, which is a weaker constraint than that coming from stability considerations.
Unless $\Lambda \ll 1$, magnetic fields of the order $\gtrsim 10^{15}$ G are needed to noticeably shift the $g_{1}$-mode frequencies for any EOS, though marginally weaker (though still strong) fields of order $\gtrsim 10^{14}$ G can significantly adjust the $g_{2}$-mode frequencies.
Given that the $g_1$-mode typically oscillates at $\sim100 \text{ Hz}$, the (rotating frame) frequency becomes negative when the ratio $\Lambda$ is less than the value (black line) that implies the maximal toroidal strength for $\Bc=10^{15}\text{ G}$, indicating the onset of instability. Moreover, the frequency shifts for overtones $(n>1)$ are less sensitive to $\Lambda$ than $g_1$-modes, resulting from nodes of displacements in the region where the toroidal component of magnetic field is non-trivial. The coupling between these modes and the structure of magnetic field is thus more tenuous. For $g_2$-modes, a toroidal-to-poloidal ratio $\Lambda$ of $\lesssim0.01$ is needed in order that $\delta\omega^B$ becomes negative; and the shifts are always positive (for stable values of $\Lambda$) for $g_3$-modes, though not shown here.
Fig.~\ref{fig:bfreqAPR} shows modified mode frequencies of $g_{1}$-modes of a specific star for the cases $\Lambda =1.0$ (top panel), $\Lambda =0.3$ (middle panel), and $\Lambda = 10^{-3}$ (bottom panels), for various values of $\Bc$, as functions of $\delta$. There $\omega_{0}$ denotes the unperturbed frequency.
As $\delta\lesssim0.01$, the absolute values of frequency corrections increase as the stratification weakens, i.e.~$\delta$ is lower, in that unperturbed frequencies in the denominator of the right hand side of \eqref{eq:modmag} converges to zero faster than the numerator.

\begin{figure} 
	\centering
	\includegraphics[scale=0.33]{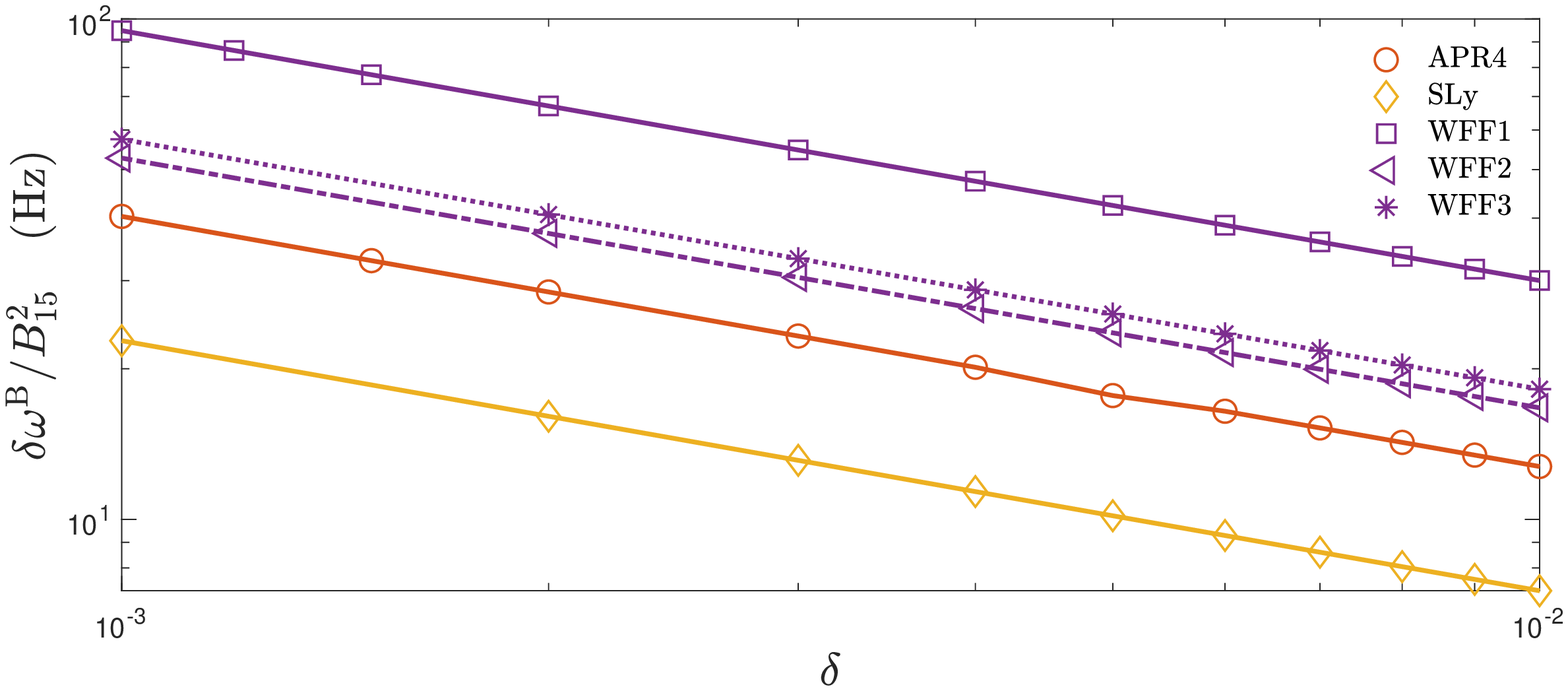}  
	\includegraphics[scale=0.33]{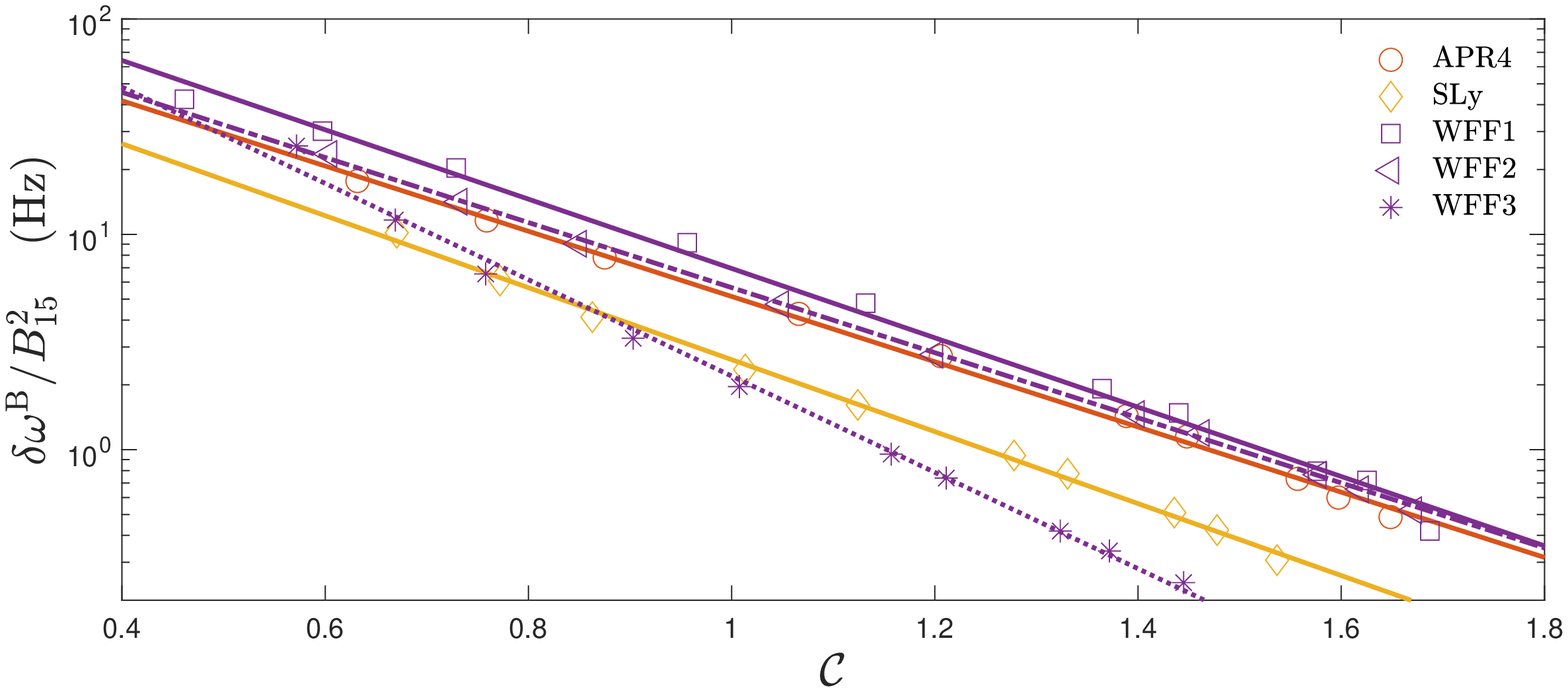}  
	\caption{Magnetic-driven modifications in eigenfrequencies of $l=2$, $g_1-$mode as functions of stratification $\delta$ (\emph{top}), and the compactness $\mathcal{C}$ (\emph{bottom}), respectively.
		The markers in the upper panel are the numerical results for the star of each EOS that has central denisty of $8\times10^{14}$ g/cm$^{3}$, while the markers in the bottom panel represent the stars described in Fig.~\ref{fig:eosmodels} with fixed $\delta=0.005$.
		We have taken $\Bc=10^{15}$ G and EOSs APR4, SLy, and WFF1-3 are included.
		In both plots, the rigid lines are the fitting results of corrections in magnetic frequencies for each EOSs.
	}
	\label{fig:deltaBfit}
\end{figure}

It is noticeable that the corrections are more severe for less compact stars when a purely poloidal ($\Lambda=1$) field is considered, as shown in Fig.~\ref{fig:deltaBfit}.
For instance, defining the compactness as $\mathcal{C}=\Mn/\Rn$, we see that for $\Bc= 10^{15}\text{ G}$, $\delta=0.005$, and EOS SLy, the correction for the $g_1$-mode is $\delta \omega^B=42.40 \text{ Hz}$ for the model with $\mathcal{C}=0.461$,
while it is $\delta \omega^B=20.32\text{ Hz}$ for the model with $\mathcal{C}=0.729$.
Additionally, we find fitting relations for the effect of magnetic field on the $g_1$-modes as 
\begin{subequations}
\begin{align}\label{eq:magshift}
	 \delta\omega^{B}  \approx \Bn^{2} e^{(c_{1}\ln\delta+c_{0})
	(d_{1}\mathcal{C} + d_{0})} \text{ Hz}.
\end{align}
\end{subequations}
The fitting coefficients for different EOS are summarized in Tab.~\ref{tab:coeff_B}.

\subsection{Tidal forces}\label{sec.V.B}

\begin{table}
	\centering
	\caption{Coefficients of the fitting functions \eqref{eq:magshift} for the magnetic-driven frequency modifications.}
	\begin{tabular}{|c|c|c|c|c|c|}
		\hline
		\quad & APR4 & SLy & WFF1 & WFF2 & WFF3 \\
		\hline
		$c_{1}$ & -0.493 & -0.492 & -0.498 & -0.494 & -0.495 \\
		\hline
		$c_{0}$ & 0.279 & -0.289 & 1.112 & 0.547 & 0.624 \\
		\hline
		$d_{1}$ & -3.489 & -3.848 & -3.707 & -3.480 & -5.150 \\
		\hline
		$d_{0}$ & 5.126 & 4.811 & 5.645 & 5.213 & 5.937 \\
		\hline
	\end{tabular}
	\label{tab:coeff_B}
\end{table}

The tidal force generated by the companion, as exerted on the primary, reads
\begin{align}
	\delta F^{T}_{\mu} =\frac{\Mc}{a^{3}}( \Po )\nabla_{\mu}(r^{2}Y_{22}).
\end{align}
The equation for the frequency shift driven by this force is found to be
\begin{align}
	\delta\omega^{T}_{\alpha}&=\frac{\Mc}{2\omega_{\alpha}a^{3}}
	\frac{  \mathlarger{\int_{\text{primary}} (\Po) \nabla_{\mu} \PT  \bar{\xi^{\mu}} \sqrt{-g}d^{3}x}  }
	{\mathlarger{\int_{\text{primary}}(\Po) e^{-2\Phi}\xi^{\mu}\bar{\xi_{\mu}}\sqrt{-g}d^{3}x}}.
\end{align}
This form is used in Eq.~\eqref{eq:modtid}. The tidal force modifies the eigenfrequencies of QNMs via the interaction mediated by the pressure (hence density) variation. Consequently, it leads to minute frequency corrections ($\sim 0.01 \%$) for $g-$modes since $g-$modes only perturb the pressure profile slightly.

\subsection{Rotation}\label{sec.V.C}

We treat the rotation of the star as a perturbation over the non-spinning equilibrium, since the Coriolis force is proportional to the square of the angular velocity, and thus a slow perturbation, to linear order, does not induce any hydromagnetic changes to the background structure \citep{Hartle67}. We also omit the spin-orbit interaction.
A (uniform) rotation $\Omega$ introduces a $g_{t\phi}$ component to the metric, causing frame dragging.
When a slow rotation is considered, this effect is small and we therefore ignore it in this work.
On top of the metric corrections, rotation also introduces the axial component
\begin{align}
	 u_{\text{rot}}^{\mu}= \Omega e^{-\Phi}\partial_{\phi},
\end{align}
to the 4-velocity, when we are working in the inertial frame.
The axial velocity $u_{\text{rot}}^{\mu}$ adds an extra term to Eq.~\eqref{Lag2vel}, resulting in
\begin{align}
	\delta u^{\mu} = i (\omega_{\alpha}+m\Omega)e^{-\Phi}\xi^{\mu},
\end{align}
and thus leads to a perturbing force
\begin{subequations}
\begin{align}
	&\delta F_{R}^{r} = 2(\Po)e^{-2\Phi} \omega_{\alpha}\Omega  (m\xi^{r}-ire^{-2\lambda}\sin^{2}\theta\xi^{\phi}), \\
	&\delta F_{R}^{\theta} = 2(\Po)e^{-2\Phi} \omega_{\alpha}\Omega (m\xi^{\theta}+i\sin\theta\cos\theta\xi^{\phi}),\\
	&\delta F_{R}^{\phi} = 2(\Po) e^{-2\Phi} \omega_{\alpha}\Omega \bigg(m\xi^{\phi}-i\frac{\xi^{r}}{r}-2i\cot\theta\xi^{\theta} \bigg).
\end{align}
\end{subequations}
Therefore, the relativistic leading order rotational corrections in the mode frequencies having the expression
\begin{align}\label{eq:modrot}
	\delta\omega^{R}_{\alpha} = -m\Omega (1-C_{nl}),
\end{align}
with 
\begin{align}
	C_{nl}=&\frac{1}{\Ms\Rs^2}\displaystyle{\int}_{\text{primary}} (\Po)e^{\Phi+\lambda}r^{2l} \times \nonumber\\
	& \big[ -e^{-\lambda}(\bar{V}_{nl}W_{nl}+\bar{W}_{nl}V_{nl})
	+V_{nl}\bar{V}_{nl} \big] dr.
\end{align}
In the Newtonian limit, this agrees with that of \cite{Unno:1979} and \cite{Strohmayer91}.

\begin{figure} 
	\centering
	\includegraphics[scale=0.35]{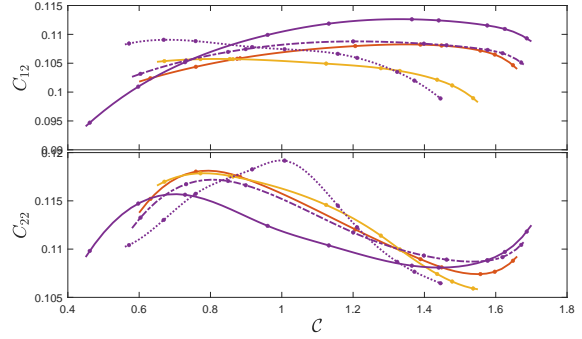}  
	\caption{Coefficients $C_{12}$ for $g_1-$modes (top panel) $C_{22}$ for $g_2-$modes as functions of compactness $\mathcal{C}$ assuming  EOS APR4 (red curves), SLy (yellow curves), and WFF1-3 (purple solid, dashed, and dotted curves, respectively).
	We fix $\delta=0.005$. Markers on each curve are models pointed out in Fig.~\ref{fig:eosmodels}. }
	\label{fig:drag}
\end{figure}

Fixing $\delta=0.005$, we plot $C_{nl}$ of $g_1-$modes ($C_{12}$, top panel) and of $g_2-$modes ($C_{22}$, bottom panel) as functions of compactness the mean density of the star in Fig.~\ref{fig:drag}.
The values for $C_{12}$ and $C_{22}$ differ only slightly \citep{Yoshida00,Passamonti09,Doneva13}, e.g.~$C_{12}=0.11$ and $C_{22}=0.112$ for the star of WFF1 EOS that has $1.4 M_{\odot}$.
On the other hand, we find that $C_{n2}$ depends only slightly on stratification $\delta$ for $n \lesssim 5$, e.g., the difference between the values of $C_{12}$ for $\delta=0.001$ and $\delta=0.01$ is $\sim0.001$ (percent level at most).
The insignificant dependence on $\delta$ of $C_{12}$ has also been shown in \cite{Gaertig09}.

\section{Crustal strain}
\label{sec.VI}
Having considered modulations in mode eigenfrequencies by tidal and magnetic fields, and the (slow) rotation of the equilibrium in Sec.~\ref{sec.V}, we now turn to investigate the maximal strain exerted on the stellar crust as a result of resonant $g-$mode displacements.

Time-varying displacements $\boldsymbol{\xi}$ between the material elements of the neutron star generates a stress. 
However, in GR, the total strain is not only due to the displacement, and there is a contribution from the perturbation of the metric to the strain tensor \citep{Carter:1972,Carter:1975,Xu:2001bu}, whose total form reads
\begin{align} \label{eq:straindefn}
	\sigma_{\mu\nu} 
	=& \frac {1} {2} \left( \nabla_{\mu} \xi_{\nu} + \nabla_{\nu} \xi_{\mu} \right)
	+\frac{1}{2}h^{\eta}_{\mu}h^{\kappa}_{\nu}\delta g_{\eta \kappa} \nonumber\\
	=& \frac {1} {2} \left( \partial_{\mu} \xi_{\nu} + \partial_{\nu} \xi_{\mu} 
	+ \delta g_{\mu\nu} \right) - \Gamma^{\sigma}{}_{\mu\nu}\xi_{\sigma},
\end{align}
where we retain just the first order terms in the second line of the equation.
Oscillations may lead to a crust failure for large enough stresses, which can be probed by the commonly used ``von Mises stress'' criterion,  coming from classical elasticity theory \citep{ll70}.
Defining the strain as \citep{Johnson13,Andersson:2018xmu}
\begin{equation} \label{eq:vonmises}
	\sigma \equiv \sqrt{ \tfrac {1} {2} \sigma_{\mu\nu} \bar{\sigma}^{\mu\nu} },
\end{equation}
then the von Mises criterion implies that the crust breaks if $\sigma$ exceeds some critical threshold, $\sm$.
In a recent semianalytic lattice stability models of \cite{Baiko18}, they calculate the threshold as $\sm \approx 0.04$ while \cite{Horowitz09} follow molecular dynamics simulations to find $\sm\approx0.1$ for low temperature stars. We adopt the former in this article with a remark that if the latter had been adopted, the amplitudes of resonantly-excited modes would need to be much higher to instigate failure.

Equation \eqref{eq:straindefn} and definition \eqref{eq:vonmises} indicate that the stress generated by the displacement $\xi_{\alpha}$ is proportional to its amplitude $q_{\alpha}$, which evolves according to equation \eqref{eq:eomosc}.
Therefore we have 
\begin{align}\label{eq:sigma(q)}
	\sigma_{\alpha}(t)= \sqrt{2}\sum_{\alpha} \sqrt{q_{\alpha}(t)\bar{q}_{\alpha}(t)} \sigma_{\alpha},
\end{align}
where $\sigma_{\alpha}$ is the \emph{unit} strain caused by $\xi_{\alpha}$ (i.e.,~for $q_{\alpha}=1$) and the pre-factor $\sqrt{2}$ is attributed to the duality of modes with $\omega$ and $-\overline{\omega}$.

\begin{figure} 
	\centering
	\includegraphics[scale=0.45]{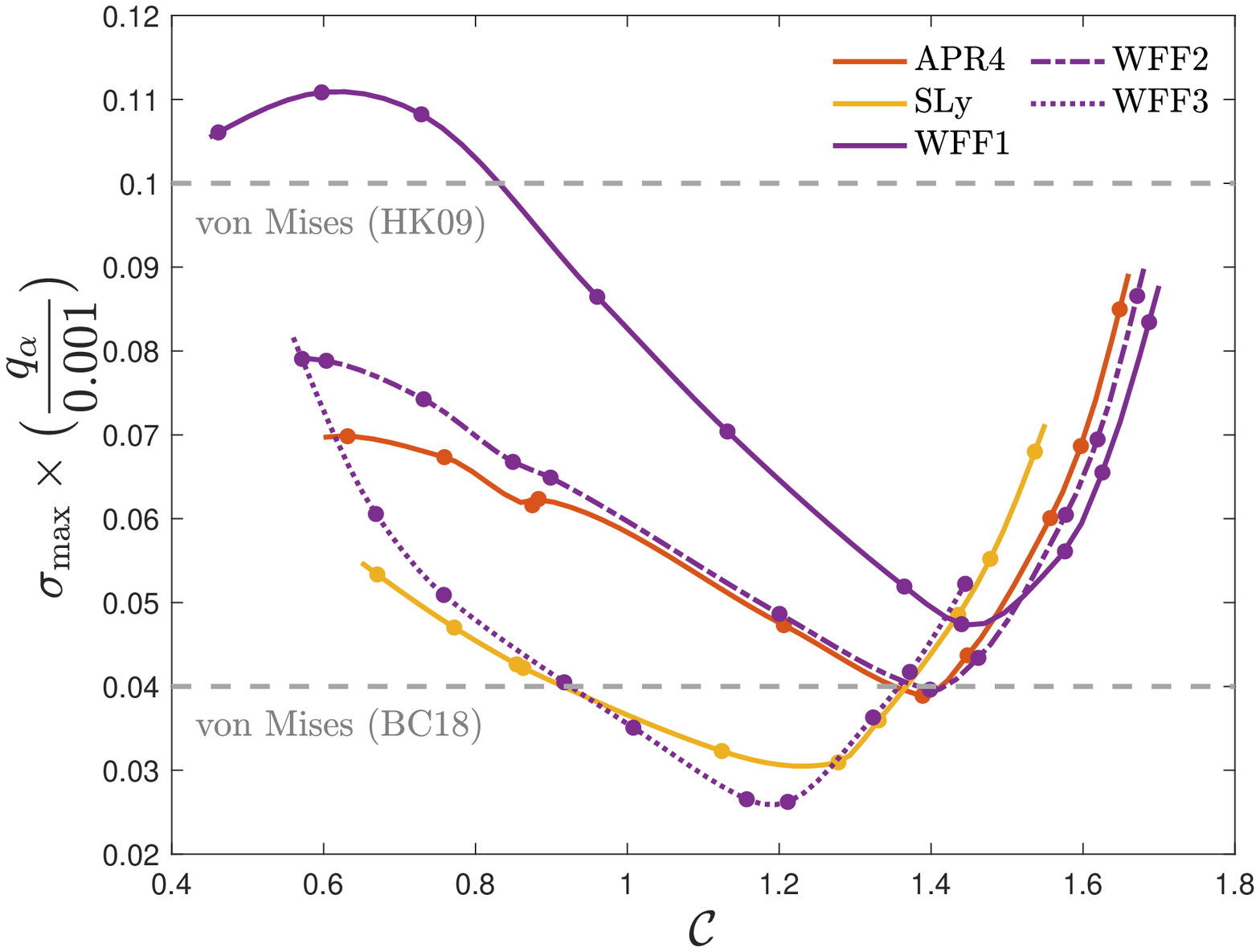}  
	\caption{Maximal crustal strain $\sigma_{\text{max}}$ due to $g_1-$modes for APR4, SLy, and WFF1-3  EOS as functions of $\mathcal{C}$. The gray dashed line represents the von Mises criterions by \protect\cite{Baiko18} and \protect\cite{Horowitz09}.
	We have taken $\delta=0.005$.}
	\label{fig:sig_cpt}
\end{figure}

Taking crust as the part of star with $0.9\Rs<r<\Rs$, in Fig.~\ref{fig:sig_cpt} we plot the maximal values of strain strain $\sigma_{\text{max}}$ in the crust due to the ($l=m=2$) $g_1-$modes for several EOS, where the stratification is taken to be $\delta=0.005$. This latter value in particular is typical in the literature for mature NSs \citep{Reisenegger:2008yk,Xu:2017hqo}.
As such, relation \eqref{eq:maxamp} implies that $g_1-$modes with tidal coupling strength $\gtrsim 8\times10^{-5}$ may be capable of generating a crustal strain that exceeds the von Mises criterion provided $\omega M\sim0.003$ for $g_1-$modes. We therefore conclude that tidal resonances in NSNS binaries can excite $g$-modes to the point that the crust may yield, which can have important implications for observations of precursors of short gamma-ray bursts \citep{Tsang:2011ad,pap1}. This latter aspect will be covered in detail in paper II.

In addition to low order $g$-modes, it has been shown by \cite{Passamonti:2020fur} that the excitation of $f$-modes before the merger, though not resonantly instigated, can generate a strain that meets the von Mises criterion. For instance, we find the strain $\sm=0.107$ for the $f$-mode of a particular primary with the SLy EOS and $M=1.27M_{\odot}$. However, only within less than 10 ms prior to the merger can $\sm$ hit the critical value of $0.04$. Though the excited $f$-modes are irrelevant to the precursors, their influences on the (phase of) GW waveforms may be measured with future GW detectors [see, e.g., \cite{Schmidt19,Pratten20}].

\section{Discussion}\label{sec.VII}

The tidal field sourced by the companion as part of a NSNS binary perturbs the primary, leading to QNM excitations. In particular, as the perturbing frequency matches to the frequencies of certain QNMs, they will be brought into resonance, during which the mode amplitudes increase rapidly \citep{Lai94b,Kokkotas:1995xe}. Though mode resonances happen all during the inspiral, those that occur in the final stages are of particular interest in that modes with higher frequency couple more strongly to the tidal field, meaning larger amplitude become available during an appropriate resonance timescale. If there is one resonantly excited mode that stresses the crust to the point beyond it cannot respond elastically, the crust may  yield. For magnetized stars, the energy released by crustal failure is likely to generate electromagnetic flares \citep{Tsang:2011ad,pap1}, which in turn offers a probe into the NS progenitor.

Our investigation on potential crust failure during the resonance between the tidal-driving frequency and the $g-$modes of a strongly magnetized primary ($\Bc\sim 10^{15} \text{G}$), is divided into three parts: 
(i) Tidal excitation increases the mode amplitude to a maximum value which depends on the coupling strength of QNMs [see the fitting Eq.~\eqref{eq:maxamp} for $l=2$, $g_{1}-$modes, where the magnetic frequency shift and stellar rotation are ignored];
(ii) Mode frequencies are modified by magnetic fields, which become insignificant for purely poloidal fields with $\Bc\lesssim 10^{15} \text{G}$ for $l=2$, $g_{1}-$modes [Fig.~\ref{fig:bfreqAPR}; though if $\Lambda \ll 1$ weaker fields can still non-trivially modulate the spectrum]; by tidal fields sourced the companion [Eq.~\eqref{eq:modtid}], which are included only for completeness since they are negligibly small ($\lesssim 0.01 \%$) for $g-$modes; and by the (uniform) rotation [Eq.~\eqref{eq:modrot}];
(iii) The maximal strain during the resonance is estimated by the maximal amplitude of QNMs [Eq.~\eqref{eq:maxamp} and Eq.~\eqref{eq:sigma(q)}].

Previous studies on tidally-driven crustal fracture use the Keplarian orbit, quadruple formula for GWs, tidal effects encoded in the Newtonian potential, and the stellar normal modes in the Newtonian theory \citep{Tsang:2011ad,Tsang:2013mca,pap1}. In our calculation, several extensions are considered to better understand the realistic applicability of the mechanism, including 3 PN orbital evolution, relativistic pulsations, and several realistic EOS that pass the constraints set by GW170817 \citep{Abbott18prl}. Our main results regarding the plausibility of $g$-modes breaking the crust are summarised in Figure \ref{fig:sig_cpt}.

Note that we have focused on $g-$mode resonances, though we include (slow) stellar rotation, which enriches the pulsation spectrum with $r-$modes. The frequencies of these latter modes are comparable with those of $g-$modes when the star rotates slowly, and thus may also be of interest in this scenario. 
In addition, our estimation on the maximal strain available in the $g-$mode resonances could be improved to better clarify the physical conditions under which crustal failure is possible.
Namely, $g-$mode resonances are influenced by various parameters, including the mass of the primary $\Ms$ and the companion $\Mc$ (or the mass ratio $q$ between them), stratification $\delta$ of the primary, rotation frequency $\nu$ of the primary, characteristic magnetic strength $\Bc$, and the poloidal-to-toroidal strength $\Lambda$.
An intensive investigation of $g-$mode resonances over a multidemensional parameter space spanned by these parameters are, therefore, important to constrain the physical conditions that allow a large enough strain that may cause crust yielding.
A forthcoming paper in this series aims to make progress in this direction.

\section*{Acknowledgements}
This work was supported by the Alexander von Humboldt Foundation, the Sandwich grant (JYP) No.~109-2927-I-007-503 by DAAD and MOST, and the DFG research Grant No. 413873357.  We thank the anonymous referee for their valuable feedback, which improved the quality of the manuscript.

\section*{Data availability statement}
Observational data used in this paper are quoted from the cited works. Data generated from computations are reported in the body of the paper. Additional data can be made available upon reasonable request.

\label{lastpage}

\end{document}